\documentclass[
 reprint,
 amsmath,amssymb,superscriptaddress,
pra,
]{revtex4-2}

\usepackage{graphicx}
\usepackage{color}
\usepackage{dcolumn}
\usepackage{bm}
\usepackage{amsthm,amsmath,amssymb}
\usepackage{siunitx}
\usepackage{subcaption}
\usepackage{IEEEtrantools}
\usepackage{tabularx}
\usepackage{makecell}
\usepackage{ragged2e}
\usepackage{threeparttable}
\graphicspath{ {./figure/} }
\allowdisplaybreaks[3]

\begin{document}

\preprint{APS/123-QED}

\title{Brownian thermal birefringent noise due to non-diagonal anisotropic photoelastic effect in multilayer coated mirrors}

\author{Yu-Pei Zhang}
\author{Shi-Xiang Yang}
\affiliation{
MOE Key Laboratory of Fundamental Physical Quantities Measurement, School of Physics, Huazhong University of Science and Technology, Wuhan 430074, People’s Republic of China
}
\author{Wen-Hai Tan}
\email{tanwh7@mail.sysu.edu.cn}
\affiliation{
MOE Key Laboratory of TianQin Mission, TianQin Research Center for Gravitational Physics \emph{\&} School of Physics and Astronomy, Frontiers Science Center for TianQin, Gravitational Wave Research Center of CNSA, Sun Yat-sen University (Zhuhai Campus), Zhuhai
519082, People’s Republic of China
}

\author{Cheng-Gang Shao}
\affiliation{
MOE Key Laboratory of Fundamental Physical Quantities Measurement, School of Physics, Huazhong University of Science and Technology, Wuhan 430074, People’s Republic of China
}

\author{Yiqiu Ma} 
\email{myqphy@hust.edu.cn}
\affiliation{
% Center for gravitational experiment, 
MOE Key Laboratory of Fundamental Physical Quantities Measurement, School of Physics, Huazhong University of Science and Technology, Wuhan 430074, People’s Republic of China
}
\author{Shan-Qing Yang}
\affiliation{
MOE Key Laboratory of TianQin Mission, TianQin Research Center for Gravitational Physics \emph{\&} School of Physics and Astronomy, Frontiers Science Center for TianQin, Gravitational Wave Research Center of CNSA, Sun Yat-sen University (Zhuhai Campus), Zhuhai
519082, People’s Republic of China
}

\begin{abstract}

% The sensitivity of current vacuum magnetic birefringence (VMB) experiments is hindered by an uncharacterized birefringent noise floor.
Thermal noise in the mirror coatings limits the accuracy of today's most optical precision measurement experiments. Unlike the more commonly discussed thermal phase noise, the crystalline coating can generate thermal birefringent noise due to its anisotropic nature.
In this study, we propose that the non-diagonal anisotropic photoelastic effect induced by the Brownian motion of mirror coating layers may contribute to this noise. Employing a standard model for the coating surface, we calculate the spectrum of the non-diagonal anisotropic Brownian photoelastic(NABP) noise to be $1.2 \times 10^{-11} p_{63} f^{-1/2}/\rm{Hz}^{1/2}$. Further experiments are warranted to validate the influence of this effect and reduce its uncertainty. Our findings highlight that for high-precision experiments involving optical resonant cavities targeting signals imprinted in optical polarizations, this noise could emerge as a limiting factor for experimental sensitivity. 
% To mitigate its impact, careful selection of coating materials with low photoelastic coefficients is recommended.
\end{abstract}

\maketitle

\section{\label{sec:sec1}Introduction}
The investigation of thermal noise in multilayer-coated mirrors is crucial for understanding its impact on the sensitivity of high-precision measurement experiments conducted within optical resonant cavities. Prominent examples include laser interferometer gravitational-wave detectors\,\cite{harry2002thermal,villar2010measurement,chalermsongsak2014broadband,gras2017audio,granata2020progress} and optical clocks\,\cite{numata2003wide,numata2004thermal,notcutt2006contribution,matei20171,ma2020investigation,yu2023excess}. While the thermal noises in these experiments have traditionally centered on the optical phase effect, extensively explored in prior studies\,\cite{levin1998internal,numata2004thermal,evans2008thermo,hong2013brownian}, we direct attention to experiments involving signals imprinted on the polarization of light, such as the measurement of vacuum magnetic birefringence (VMB)\,\cite{cameron1993search,ejlli2020pvlas,fan2017oval,agil2022vacuum,chen2007q}, or the search for axion-like particles\,\cite{ehret2010new,liu2019searching,obata2018optical}, where thermal birefringent noise emerges as a key factor.

Low-loss mirrors are constructed by sequentially depositing two different materials onto the substrate, resulting in thermal noise contributed by the coating, the substrate, and by their interface. Here the coating is the main contributor because that optical field is circulated within the cavity by reflecting from the coating layers. Coating thermal noise has two potential sources: (1) temperature variations due to thermal dissipation and (2) Brownian motion due to mechanical dissipation. Due to their low mechanical losses, crystalline coatings have lower Brownian noise than amorphous coatings \cite{cole2013tenfold,chalermsongsak2016coherent}. However, due to the anisotropic nature of the crystalline coating, its optical properties may exhibit an \textit{anisotropy} which is different from that of the amorphous coatings.

Recent investigations into polarization fluctuations resulting from temperature fluctuations and the \textit{anisotropic} thermal properties of crystalline coating materials, so called thermorefringent noise, have been reported\,\cite{kryhin2023thermorefringent}. Previous studies\,\cite{kondratiev2011thermal, harry2012optical} have considered the photoelastic effect due to Brownian motion in the coating layers, but included only the \textit{isotropic} part which is the diagonal terms in the photoelastic coefficients matrix\,(i.e, $p_{ij}=p\delta_{ij}$). The optical \textit{anisotropy} produced by the photoelastic effect can be divided into two types. The first one is due to the the unequal diagonal elements which produces the birefringence, and this effect has been studied\,\cite{ejlli2020pvlas} previously. The other one is due to the  the non-diagonal elements, which leads to the variation in the angle of birefringence\,(see Section\,\ref{sec:sec2A}), which remains unexplored. In this work, we will explore the possible Brownian noise contributed by the non-zero off-diagonal photoelastic coefficients, which could result from the strains induced during the manufacturing process. 

Brownian motion of the coating layer, through the non-diagonal anisotropic photoelastic effect, can lead to fluctuations in the polarization state. This noise will appear in the polarization mode that is perpendicular to the incident polarization, termed in this paper as
non-diagonal anisotropic Brownian photoelastic (NABP) noise. This noise has the potential to limit the performance of VMB experiments and other precision optical polarization measurement experiments. Therefore, experimental characterisation of the off-diagonal photoelastic coefficient of the coating materials is necessary for an accurate assessment of the NABP noise.

This paper is structured as follows: In Section\,\ref{sec:sec2}, we begin by introducing the theoretical model of the photoelastic effect and its impact when light traverses an anisotropic photoelastic medium. Subsequently, we give the expression for the polarization fluctuations of the reflected light employing the standard structure of mirror coating. In Section\,\ref{sec:sec3}, the noise level of the thermal birefringent noise contributed by the non-diagonal anisotropic photoelastic effect is calculated  and compared with other noise. Finally, Section\,\ref{sec:sec4} summarizes our principal findings and engages in a about future avenues of research.

\section{\label{sec:sec2}Theoretical Model}
\subsection{\label{sec:sec2A}Photoelastic effect}
\begin{figure}[t]
\includegraphics[width=8.6cm]{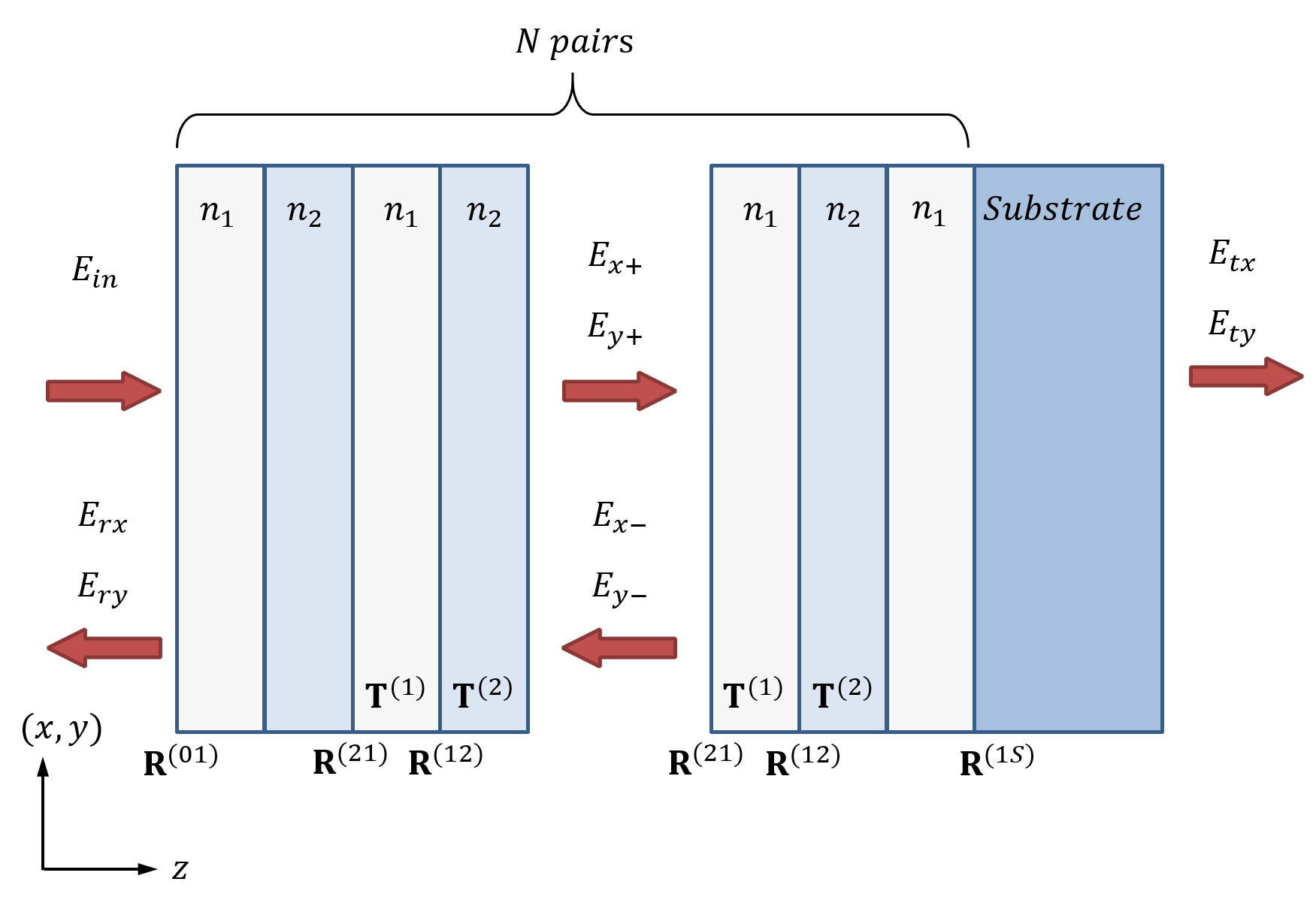}
\captionsetup{font={footnotesize}, justification=RaggedRight}
\caption{Structure of multilayer coated mirror. Light colored part represents coating layer, dark colored part represents substrate. The subscript 1 represents $\rm{GaAs}$ layers and 2 represents $\rm{AlGaAs}$ layers in our experiments. The thickness of all coating layer is $\lambda /4$ except the first one, which is $\lambda /2$.}
\label{fig:Coating structure}
\end{figure}

The photoelastic effect, denoting the change in refractive index induced by mechanical strain, is characterized by symmetric tensors representing the strain ($u_{ij}$) and the change in the optical indicatrix ($\Delta B_{ij}$). Utilizing contracted indices ($i, j \in {xx, yy, zz, yz, zx, xy}$ rather than $x,y,z$) for these tensors, the photoelastic effect can be expressed in a vectorial form:
\begin{IEEEeqnarray}{rCl}
\Delta B_{i}= p_{ij}u_{j},
\end{IEEEeqnarray}
where $\Delta B_{i}$ is the change of the optical indicatrix, $p_{ij}$ is photoelastic tensor, $u_{j}$ is strain components\,(we follow the conventions used in\,\cite{yariv1983optical} where the six independent matrix elements of $u_{ij}$ is summarized as one vector $u_j$, the same for the relationship between the $\Delta B_{ij}$ and the $\Delta B_i$). Since fluctuations caused by photoelastic effect in the transverse direction are much smaller than in the longitudinal direction\,\cite{hong2013brownian}, we only consider the variation of the refractive index caused by the change of coating thickness
\begin{IEEEeqnarray}{rCl}
\Delta B _{i}= p_{i3}\delta d/d,
\end{IEEEeqnarray}
where $\delta d/d$ denotes the change in coating thickness.
Since light propagates along the z-axis (see Fig.\ref{fig:Coating structure}), and the optical field is confined to the x-y plane, the optical indicatrix $\mathbf{B}$ is given by:
\begin{IEEEeqnarray}{rCl}
\mathbf{B} =\begin{bmatrix}B_{xx} &0 \\ 0&B_{yy}\end{bmatrix}+\begin{bmatrix}p_{13} & p_{63}\\p_{63}&p_{23}\end{bmatrix}\frac{\delta d}{d},
\end{IEEEeqnarray}
while the dielectric tensor $\boldsymbol{{\rm \varepsilon}}$ is inverse of $\mathbf{B}$. To the first order of $\delta d/d$, the dielectric tensor reads:
\begin{IEEEeqnarray}{rCl}
&\boldsymbol{{\rm \varepsilon}}& = \mathbf{B}^{-1}\\ 
&=& \begin{bmatrix}\left \langle n_{x}^{2} \right \rangle & 0\\0 &\left \langle n_{y}^{2} \right \rangle \end{bmatrix}
-\begin{bmatrix}p_{13}\left \langle n_{x}^{4} \right \rangle  & p_{63}\left \langle n_{x}^{2} \right \rangle \left \langle n_{y}^{2} \right \rangle\\p_{63} \left \langle n_{x}^{2} \right \rangle \left \langle n_{y}^{2} \right \rangle &p_{23}\left \langle n_{y}^{4} \right \rangle\end{bmatrix}\frac{\delta d}{d},\nonumber
\end{IEEEeqnarray}
where $\left \langle n_{x} \right\rangle ,\left\langle n_{x} \right \rangle$ is refractive index in $x,y$ direction, $\left \langle n_{x} \right \rangle={B}_{xx}^{-1/2},\left \langle n_{y} \right \rangle={B}_{yy}^{-1/2}$. The strain variation  $\delta d/d$ changes the principle directions of the dielectric tensor,  where the refractive index and the new principle directions are given by the diagonalisation of $\boldsymbol{{\rm \varepsilon}}$:
\begin{IEEEeqnarray}{rCl}
n_{x}&\approx&\left \langle n_{x} \right \rangle- \frac{p_{13}\left \langle n_{x}^{3} \right \rangle}{2} \frac{\delta d}{d},
\label{eq:nx}
\\
n_{y}&\approx&\left \langle n_{y} \right \rangle- \frac{p_{23}\left \langle n_{y}^{3} \right \rangle}{2} \frac{\delta d}{d},
\label{eq:ny}
\end{IEEEeqnarray}
with corresponding eigenvectors
\begin{IEEEeqnarray}{rCl}
&\mathbf{v}_{1}&\approx\begin{bmatrix}1\\0\end{bmatrix}- \frac{p_{63}\left \langle n_{x}^{2} \right \rangle \left \langle n_{y}^{2} \right \rangle}{\left \langle n_{x}^{2} \right \rangle - \left \langle n_{y}^{2} \right \rangle} \frac{\delta d}{d}\begin{bmatrix}0\\1\end{bmatrix},
\label{eq:vector1}
\\ &\mathbf{v}_{2}&\approx\begin{bmatrix}0\\1\end{bmatrix}+ \frac{p_{63}\left \langle n_{x}^{2} \right \rangle \left \langle n_{y}^{2} \right \rangle}{\left \langle n_{x}^{2} \right \rangle - \left \langle n_{y}^{2} \right \rangle} \frac{\delta d}{d} \begin{bmatrix}1\\0\end{bmatrix}.
\label{eq:vector2}
\end{IEEEeqnarray}
This fluctuations in eigenvectors is equivalent to intrinsic birefringence of coating rotated by a small angle 
\begin{IEEEeqnarray}{rCl}
\delta \theta \approx -\frac{p_{63}\left \langle n_{x}^{2} \right \rangle \left \langle n_{y}^{2} \right \rangle}{\left \langle n_{x}^{2} \right \rangle - \left \langle n_{y}^{2} \right \rangle} \frac{\delta d}{d}.
\label{eq:polarizaton angle change}
\end{IEEEeqnarray}
Notably, while the influence of $p_{13}$ and $p_{23}$ on fluctuations in $n_{x}$ and $n_{y}$ has been previously studied\,\cite{kondratiev2011thermal}, the impact of the $p_{63}$ term, leading to fluctuations in the angle of intrinsic birefringence, remains unexplored. 

Typical optical materials used in the crystalline coating are GaAs/AlGaAs. Theoretically both materials are isotropic and therefore do not have non-diagonal photoelastic coefficients $p_{63}$. Here we propose two possible physical scenarios for producing a non-zero $p_{63}$. Since the thermal expansion coefficients of the coating $\alpha_{C}$ and the substrate $\alpha_{S}$ are different, the thermal elasticity will induce a strain in the coating layers when the temperature drops from the manufacturing temperature to the room temperature\, \cite{vyatchanin2020loss}:
\begin{equation}
    \frac{\delta l}{l}=(\alpha_{C}-\alpha_{S})\Delta T=2\times10^{-3}
\end{equation}
Theoretically, the coating film is stressed uniformly, which means that 
the lattices on the coating surface are all exposed to the same stresses, and the direction of the stress is along their location to the centre of the mirror. For most of the lattices on the coating surface, the direction of the crystal axis and the stress are different. This lowers the symmetry of the crystal system from cubic to monoclinic or triclinic \cite{sun2009strain}, thus producing a non-zero $p_{63}$.

Another possible mechanism is that the deformation of the top layer is less than that of the layer closest to the substrate, creating a shear strain at the area where the light spot stays:
\begin{equation}
    S=\frac{\delta l_{t}-\delta l_{s}}{l}\frac{w_{0}}{d}
\end{equation}
where $w_{0}$ is the beam radius, $d$ is the thickness of the full coating layer, and $\delta l_{t}/l,\delta l_{s}/l$ are the strains in the top and substrate layers, respectively. This shear strain could lower the symmetry of the crystal system from cubic to triclinic \cite{sun2009strain}, thus producing a non-zero $p_{63}$. The value of $p_{63}$ needs to be measured by further experiments, and in this paper we temporarily use 1$\%$ of the photoelastic coefficient $p_{13}$ to demonstrate the influence of this effect.

\subsection{\label{sec:sec2B}Light penetration in muti-layer coating}
The conventional structure of a low-loss mirror coating, depicted in Fig.,\ref{fig:Coating structure}, consists of $N$ pairs of layers with high and low refractive index. The thickness of most coating layers is $\lambda /4$, except for the top layer which has a thickness of $\lambda /2$. To compute the light field reflected from the mirror coating, we adopt a method similar to that employed by LIGO in calculating thermorefringent noise,\cite{kryhin2023thermorefringent}. Commencing with transfer matrices for a unit cell, we deduce the matrix representing the fluctuations in the optical field reflected from the mirror coating.

\subsubsection{Transfer Matrix of a unit cell}
As mentioned above, mirror coating consists of repeating pairs of high and low refractive coating index layers, we denote the propagation matrix of the $i_\text{th}$ layer pair as a unit cell propagation matrix $\mathbf{\Phi}_{i}$. A unit cell is composed of two layers and two interfaces between them. A 4-dimensional vector is used to represent the light propagating through the coating:
\begin{IEEEeqnarray}{rCl}
\mathbf{E}=\begin{bmatrix}
 E_{x+}\\E_{x-}\\E_{y+}\\E_{y-}
\end{bmatrix},
\end{IEEEeqnarray}
where $x,y$ is the direction of polarization, $\pm$ represent the (right/left) direction of light propagation.

The propagation matrix in the bulk medium is
\begin{IEEEeqnarray}{rCl}
\mathbf{T}^{(I)}=\begin{bmatrix}
\mathbf{T}^{(I)}_{x}&0\\0&\mathbf{T}^{(I)}_{y}
\end{bmatrix},
\label{eq:T Definition}
\end{IEEEeqnarray}
where superscript $I$ denoted the material\,($I=1/2$ represents the ${\rm SiO}_2$ and ${\rm Ta}_2{\rm O}_5$). $\mathbf{T}^{(I)}_{x}$ and $\mathbf{T}^{(I)}_{y}$ are the propagating matrix of the electric field $E_{x/y}$ with the coordinate system using basis vector $\mathbf{v}_{1},\mathbf{v}_{2}$ (in \cite{kryhin2023thermorefringent}, these two vectors are defined as ``coordinate vectors"), which is the same in Eqs.(\ref{eq:vector1}) and (\ref{eq:vector2}), $\mathbf{T}^{(I)}_{a}$ is 
\begin{IEEEeqnarray}{rCl}
\mathbf{T}^{(I)}_{a}=\begin{bmatrix}
e^{-in^{(I)}_{a}kd^{(I)}_{i}}&0\\0&e^{in^{(I)}_{a}kd^{(I)}_{i}}
\end{bmatrix},
\label{eq:Tx Definition}
\end{IEEEeqnarray}
where $n^{(I)}_{a}$ is the refractive index in the material $I$ along the $a$-direction, $k$ is the wave vector of the incident light, $d^{(I)}_{i}$ is the thickness of the material $I$ in $i$-th layer. 
Since the coordinate vectors of a layer fluctuate by an angle $\delta\theta$, which is obtained by Eq.(\ref{eq:polarizaton angle change}):
\begin{IEEEeqnarray}{rCl}
\delta \theta =
\begin{cases}
-\frac{p_{63}^{(1)}n_{1}^{4}}{n_{1}^{2}-(n_{1}+\Delta n_{1})^{2}} \frac{\delta d^{(1)}_{i}}{d^{(1)}_{i}},  & \delta d^{(I)}_{i}=\delta d^{(1)}_{i}
\\[5pt] \frac{p_{63}^{(2)}n_{2}^{4}}{n_{2}^{2}-(n_{2}+\Delta n_{2})^{2}} \frac{\delta d^{(2)}_{i}}{d^{(2)}_{i}},  & \delta d^{(I)}_{i}=\delta d^{(2)}_{i}
\label{eq:delta theta Definition}
\end{cases}.
\end{IEEEeqnarray}
where the $\delta d^{(I)}_{i}$ represent the deformation of the material $I$ layer in the $i$-th unit cell.  Here we denote the optical anisoptropy using the difference of refractive index along x/y directions $n_{y}^{I}=n_{x}^{I}+\Delta n_{I}$ and we have  $\Delta n_{I}/n_x^I \ll 1$.  The $\mathbf{R}^{(IJ)}$ is the transfer matrix of the interface between material $I$ and $J$, which is combined by a rotation matrix and the transfer matrix:
\begin{IEEEeqnarray}{rCl}
&\mathbf{R}^{(12)}&=\begin{bmatrix}
\mathbf{r}^{(12)}_{xx}\rm{cos}(\delta \theta)&-\mathbf{r}^{(12)}_{xy}\rm{sin}(\delta \theta)
\\[5pt] \mathbf{r}^{(12)}_{yx}\rm{sin}(\delta \theta)&\mathbf{r}^{(12)}_{yy}\rm{cos}(\delta \theta)
\end{bmatrix},
\\[5pt]
&\mathbf{R}^{(21)}&=\begin{bmatrix}
\mathbf{r}^{(21)}_{xx}\rm{cos}(\delta \theta)&\mathbf{r}^{(21)}_{xy}\rm{sin}(\delta \theta)
\\[5pt] -\mathbf{r}^{(21)}_{yx}\rm{sin}(\delta \theta)&\mathbf{r}^{(21)}_{yy}\rm{cos}(\delta \theta)
\end{bmatrix},
\end{IEEEeqnarray}
where $\mathbf{r}^{(IJ)}_{ab}$ is transfer matrix of interface without considering the polarization(both sides in same polarization direction): 
\begin{IEEEeqnarray}{rCl}
\mathbf{r}^{(IJ)}_{ab}=\dfrac{1}{2}\begin{bmatrix}
1+\dfrac{n^{(J)}_{b}}{n^{(I)}_{a}}&1-\dfrac{n^{(J)}_{b}}{n^{(I)}_{a}}
\\[10pt] 1-\dfrac{n^{(J)}_{b}}{n^{(I)}_{a}}&1+\dfrac{n^{(J)}_{b}}{n^{(I)}_{a}}
\label{eq:boundary condition}
\end{bmatrix},
\end{IEEEeqnarray}
which is described by the boundary conditions derived from Maxwell's equation\,\cite{kryhin2023thermorefringent}.

\begin{figure}[t]
\includegraphics[width=8.6cm]{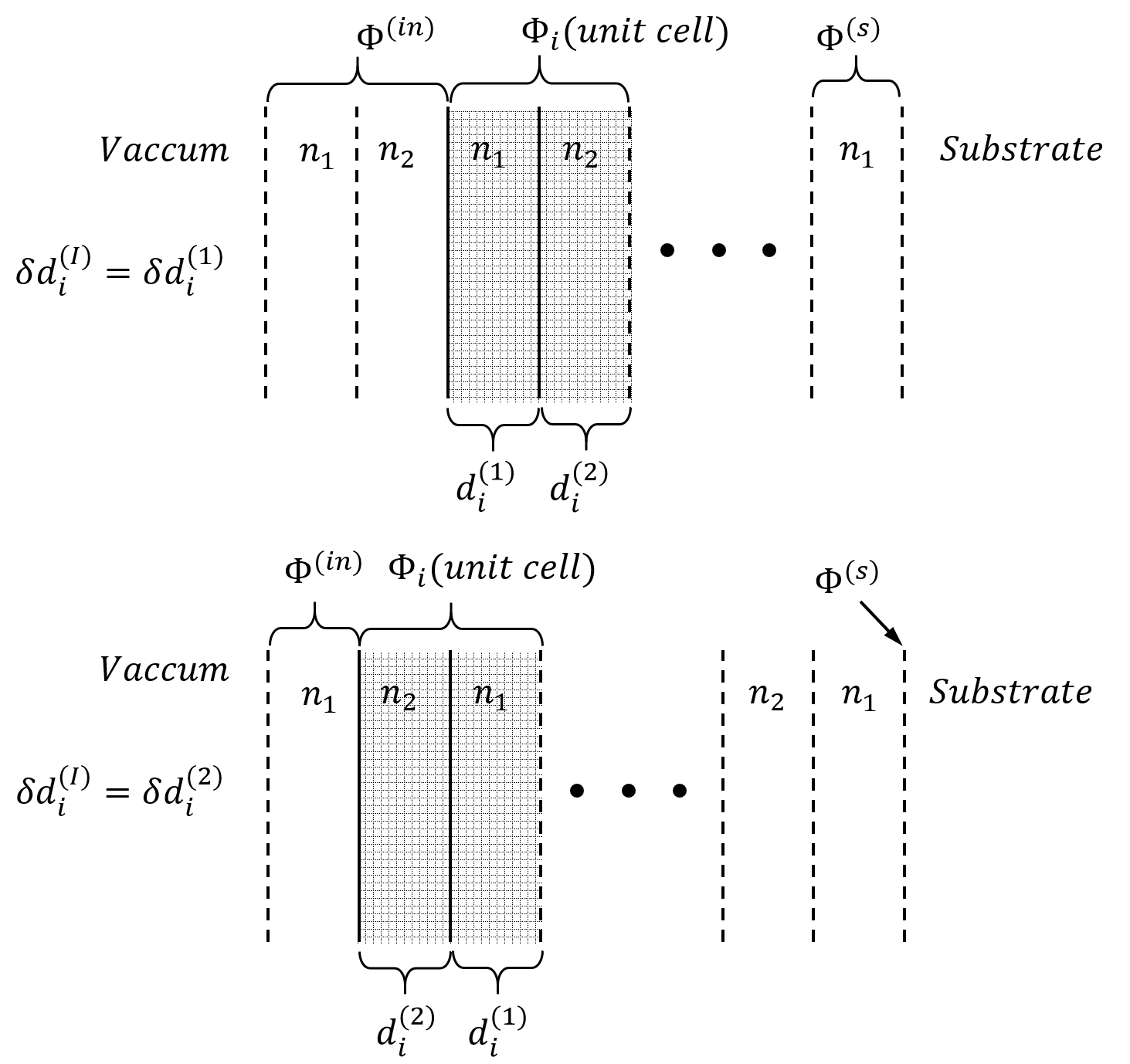}
\captionsetup{font={footnotesize}, justification=RaggedRight}
\caption{Definition of the unit cell, the incident cell and the substrate cell when calculating different material coating layer. The unit cell is marked with shadows and solid lines.}
\label{fig:Unit cell definition}
\end{figure}

As we shall see in the next section (Eq.\,\eqref{eq:Phi_i}\,\eqref{eq:M matrix expand}). To the first order of the strain deformation, the contributions of different layers to the propagation matrix are independent of each other, and the strain-modification of the full propagation matrix can be obtained by adding the contributions of all layers. Therefore, the key component of our calculation is the strain-modification of a single layer.  In the upper panel of schematic diagram of coating cell Fig.\ref{fig:Unit cell definition}, the strain deformation $\delta d_i$ of the layer $d^{(1)}_{i}$ can influence the interfaces with two neighouring layers marked as solid line, and they are both included in the unit cell $\mathbf{\Phi}_{i}$. However, for the layer $d^{(2)}_{i}$, only one interface is included in the unit cell which means that another definition is needed to calculate the influence of $\delta d^{(2)}_{i}$. Therefore, it is convenient to use two different definitions for calculating $\delta d^{(1)}_{i}$ and $\delta d^{(2)}_{i}$ :
\begin{IEEEeqnarray}{rCl}
\mathbf{\Phi}_{i}=
\begin{cases}
\mathbf{R}^{(21)}\mathbf{T}^{(1)}\mathbf{R}^{(12)}\mathbf{T}^{(2)}& \text{when } \delta d^{(I)}_{i}=\delta d^{(1)}_{i}
\\ \mathbf{R}^{(12)}\mathbf{T}^{(2)}\mathbf{R}^{(21)}\mathbf{T}^{(1)}& \text{when }\delta d^{(I)}_{i}=\delta d^{(2)}_{i}
\label{eq:Phi Definition}
\end{cases}.
\end{IEEEeqnarray}
The choice of the definition of $\mathbf{\Phi}_{i}$ depends on the material $I$ of the layer being calculated, which is presented in detail in next section (Section \ref{sec:sec2B2}).

\subsubsection{\label{sec:sec2B2}Transfer Matrix of full coating layers}

The matrix of the full coating layers can be written as the product of unit cell matrices: 
\begin{IEEEeqnarray}{rCl}
\mathbf{M}=\mathbf{\Phi}^{(in)}\left(\prod_{i=2}^{N} \mathbf{\Phi}_{i}\right)\mathbf{\Phi}^{(s)},
\label{eq:M matrix}
\end{IEEEeqnarray}
where $\mathbf{\Phi}_{i}$ represent the matrix of $i$-th unit cell, $\mathbf{\Phi}^{(in,s)}$ is the propagation matrix of the incident cell and the last coating cell near the substrate, $N$ represents the number of unit cells in the coating. 
Separating a unit cell matrix into the steady part and the fluctuating part:
\begin{IEEEeqnarray}{rCl}
\mathbf{\Phi}_{i}=\mathbf{\Phi}_{0}+\mathbf{\Phi}'\delta d^{(I)}_{i}.
\label{eq:Phi_i}
\end{IEEEeqnarray}
Since $\delta d^{(I)}_{i}$ of different layers are independent of each other\,\cite{hong2013brownian}, their influence on $\mathbf{M}$ can be calculated separately (except for the first and last layer):
\begin{IEEEeqnarray}{rCl}
\mathbf{M}(\delta d^{(I)}_{i})=
\mathbf{\Phi}^{(in)}_{0}\mathbf{\Phi}_{0}^{i-1}(\mathbf{\Phi}_{0}+\mathbf{\Phi}'\delta d^{(I)}_{i})\mathbf{\Phi}_{0}^{Q}\mathbf{\Phi}^{(s)}_{0},
\label{eq:M matrix expand}
\end{IEEEeqnarray}
where 
\begin{equation}
    Q=\begin{cases}
        N-i-2, & \delta d^{(I)}_{i}=\delta d^{(1)}_{i}
        \\N-i-1, & \delta d^{(I)}_{i}=\delta d^{(2)}_{i}
    \end{cases}
\end{equation}
The definition choice of $\mathbf{\Phi}^{(in)},\mathbf{\Phi}^{(s)}$ and $\mathbf{\Phi}_{i}$ depends on the material of the fluctuating layer $\delta d^{(I)}_{i}$ that need to be calculated. For the fluctuation of the top layer $\delta d^{(1)}_{1}$ and the layer next to the substrate $\delta d^{(1)}_{N+1}$, the $\mathbf{M}$ is:
\begin{IEEEeqnarray}{rCl}
&\mathbf{M}&(\delta d^{(1)}_{1})=(\mathbf{\Phi}^{(in)}_{0}+\mathbf{\Phi}'^{(in)}\delta d^{(1)}_{1})\mathbf{\Phi}_{0}^{N-1}\mathbf{\Phi}^{(s)}_{0},
\label{eq:M expand d1}
\\[5pt]
&\mathbf{M}&(\delta d^{(1)}_{N+1})=\mathbf{\Phi}^{(in)}_{0}\mathbf{\Phi}_{0}^{N-1}(\mathbf{\Phi}^{(s)}_{0}+\mathbf{\Phi}'^{(s)}\delta d^{(1)}_{N+1}).
\label{eq:M expand d2}
\\ \nonumber
\end{IEEEeqnarray}
Using the definition in Eqs.(\ref{eq:T Definition}-\ref{eq:Phi Definition}), it turns out that $\mathbf{\Phi}_{0}$ is block-diagonal and $\mathbf{\Phi}'$ is off-diagonal (to the first order of $\delta d^{(I)}_{i}$):
\begin{IEEEeqnarray}{rCl}
\mathbf{\Phi}_{i} = \begin{bmatrix}
\mathbf{\Phi}_{xx} & 0 \\ 0 & \mathbf{\Phi}_{yy}
\end{bmatrix}
+\begin{bmatrix}
0 & \mathbf{\Phi}_{xy} \\ \mathbf{\Phi}_{yx} & 0
\end{bmatrix}\delta d^{(I)}_{i},
\label{eq:phi0}
\end{IEEEeqnarray}
where $\mathbf{\Phi}_{xx}$ and $\mathbf{\Phi}_{xy}$ is
\begin{IEEEeqnarray}{rCl}
&\mathbf{\Phi}_{xx}& = \mathbf{\Phi}_{yy}
\label{eq:Phi xx}
\\ &\approx&
\begin{cases}
-\frac{1}{2n_{1}n_{2}}
\begin{bmatrix}
n_{1}^{2}+n_{2}^{2} & n_{2}^{2}-n_{1}^{2} 
\\ n_{2}^{2}-n_{1}^{2} & n_{1}^{2}+n_{2}^{2}
\end{bmatrix}, & \delta d^{(I)}_{i}=\delta d^{(1)}_{i}
 \\[10pt]
-\frac{1}{2n_{1}n_{2}}
\begin{bmatrix}
n_{1}^{2}+n_{2}^{2} & n_{1}^{2}-n_{2}^{2} 
\\ n_{1}^{2}-n_{2}^{2} & n_{1}^{2}+n_{2}^{2}
\end{bmatrix}, & \delta d^{(I)}_{i}=\delta d^{(2)}_{i}
\end{cases}, \nonumber
\end{IEEEeqnarray}
\begin{widetext}
\begin{IEEEeqnarray}{rcl}
\mathbf{\Phi}_{xy} = \mathbf{\Phi}_{yx} \approx
\begin{cases}
\frac{p_{63}^{(1)}n_{1}}{4 n_{2} d^{(1)}_{i}}
\begin{bmatrix}
n_{2}^{2}-n_{1}^{2}+i\pi n_{1}n_{2} & n_{1}^{2}+n_{2}^{2} 
\\ n_{1}^{2}+n_{2}^{2} & n_{2}^{2}-n_{1}^{2}-i\pi n_{1}n_{2}
\end{bmatrix}, & \delta d^{(I)}_{i}=\delta d^{(1)}_{i}
\\[10pt]
\frac{p_{63}^{(2)}n_{2}}{4 n_{1} d^{(2)}_{i} } 
\begin{bmatrix}
n_{1}^{2}-n_{2}^{2}+i\pi n_{1}n_{2} & n_{1}^{2}+n_{2}^{2} 
\\ n_{1}^{2}+n_{2}^{2} & n_{1}^{2}-n_{2}^{2}-i\pi n_{1}n_{2}
\end{bmatrix}, & \delta d^{(I)}_{i}=\delta d^{(2)}_{i}
\end{cases}.
\label{eq:phi xy}
\end{IEEEeqnarray}
\end{widetext}
The definition of the incident cell $\mathbf{\Phi}^{(in)}$ and the substrate cell $\mathbf{\Phi}^{(s)}$ are also different when calculating different material layers:

\begin{IEEEeqnarray}{rCl}
&\mathbf{\Phi}^{(in)}&=
\begin{cases}
\mathbf{R}^{(01)}\mathbf{T}^{(1)}\mathbf{R}^{(12)}\mathbf{T}^{(2)}, & \delta d^{(I)}_{i}=\delta d^{(1)}_{1}
\\ \mathbf{R}^{(01)}\mathbf{T}^{(1)}, & \delta d^{(I)}_{i}=\delta d^{(2)}_{1}
\end{cases},
\label{eq:Phi(in) Definition}
\\
&\mathbf{\Phi}^{(s)}&=
\begin{cases}
\mathbf{R}^{(21)}\mathbf{T}^{(1)}\mathbf{R}^{(1s)},  & \delta d^{(I)}_{i}=\delta d^{(1)}_{N+1}
\\ \mathbf{R}^{(1s)}, & \delta d^{(I)}_{i}=\delta d^{(2)}_{N+1}
\end{cases}.
\label{eq:Phi(s) Definition}
\end{IEEEeqnarray}
Assuming that incident light comes from the vacuum where $ n_{0}=1$, the matrix of the incident cell and the substrate cell is:
\begin{IEEEeqnarray}{rcl}
&\mathbf{\Phi}^{(in)}& =
\begin{bmatrix}
\mathbf{\Phi}_{xx}^{(in)} & \mathbf{\Phi}_{xy}^{(in)} \delta d^{(1)}_{1}
\\ \mathbf{\Phi}_{yx}^{(in)}\delta d^{(1)}_{1} & \mathbf{\Phi}_{yy}^{(in)}
\end{bmatrix},
\end{IEEEeqnarray}

\begin{equation}
\mathbf{\Phi}^{(s)}= 
\begin{bmatrix}
\mathbf{\Phi}_{xx}^{(s)} & \mathbf{\Phi'}_{xy}^{(s)} \delta d^{(I)}_{N+1}\\
\mathbf{\Phi}_{yx}^{(s)} \delta d^{(I)}_{N+1} & \mathbf{\Phi}_{yy}^{(s)}
\end{bmatrix} ,
\end{equation}
in which

\begin{IEEEeqnarray}{rcl}
\label{eq:phi in xy}
&\mathbf{\Phi}^{(in)}_{xx} &= \mathbf{\Phi}^{(in)}_{yy}
\\
&\approx&\begin{cases}
-\frac{1}{2n_{1}}
\begin{bmatrix}
n_{1}^{2}+n_{2}^{2} & n_{2}^{2}-n_{1}^{2} \\ n_{2}^{2}-n_{1}^{2} & n_{1}^{2}+n_{2}^{2}
\end{bmatrix}, & \delta d^{(I)}_{i}=\delta d^{(1)}_{i}
\\[10pt]
\frac{i}{2}\begin{bmatrix}
-1-n_{1} & 1-n_{1} \\ -1+n_{1} & 1+n_{1}
\end{bmatrix}, & \delta d^{(I)}_{i}=\delta d^{(2)}_{i}
\end{cases},\nonumber
\end{IEEEeqnarray}
and
\begin{IEEEeqnarray}{rcl}
&\mathbf{\Phi}^{(s)}_{xx}&=\mathbf{\Phi}^{(s)}_{yy}\\
&\approx &\begin{cases}
\frac{i}{2n_{1}n_{2}}\begin{bmatrix}
-n_{1}^{2}-n_{2}n_{s} & -n_{1}^{2}+n_{2}n_{s} 
\\ n_{1}^{2}-n_{2}n_{s} & n_{1}^{2}+n_{2}n_{s}
\end{bmatrix},
 & \delta d^{(I)}_{i}=\delta d^{(1)}_{N+1}
\\[10pt]
\frac{1}{2}\begin{bmatrix}
1+n_{s}/n_{1} & 1-n_{s}/n_{1} \\ 1-n_{s}/n_{1} & 1+n_{s}/n_{1}
\end{bmatrix},
 & \delta d^{(I)}_{i}=\delta d^{(2)}_{N+1}
\end{cases}, \nonumber
\end{IEEEeqnarray}
\begin{widetext}
\begin{IEEEeqnarray}{rcl}
\mathbf{\Phi}^{(in)}_{xy}&=&\mathbf{\Phi}^{(in)}_{yx} 
\label{eq:Phi in xy}\approx \begin{cases}
\frac{ p_{63}^{(1)} n_{1} }{4 d_{1}^{(1)}}
\begin{bmatrix}
n_{1}^{2}-n_{2}-\frac{i\pi}{2}n_{1}(n_{2}+1) 
& -n_{1}^{2}-n_{2}-\frac{i\pi}{2}n_{1}(n_{2}-1)
\\ -n_{1}^{2}-n_{2}+\frac{i\pi}{2}n_{1}(n_{2}-1) 
& n_{1}^{2}-n_{2}+\frac{i\pi}{2}n_{1}(n_{2}+1)
\end{bmatrix}, & \delta d^{(I)}_{i}=\delta d^{(1)}_{1}
\\ 0, & \delta d^{(I)}_{i}=\delta d^{(2)}_{1} 
\end{cases},\\
\mathbf{\Phi}^{(s)}_{xy} &=&\mathbf{\Phi}^{(s)}_{yx}
\approx \begin{cases}
\frac{p_{63}^{(1)}n_{1}}{8 d^{(1)}_{N+1} n_{2}} 
\begin{bmatrix}
-2 i (n_{1}^{2}-n_{2}n_{s})-n_{1}(n_{2}+n_{s}) \pi 
& -2 i (n_{1}^{2}+n_{2}n_{s})+n_{1}(n_{s}-n_{2})\pi 
\\ 2 i (n_{1}^{2}+n_{2}n_{s})+n_{1}(n_{s}-n_{2})\pi 
& 2 i (n_{1}^{2}-n_{2}n_{s})-n_{1}(n_{2}+n_{s})\pi
\end{bmatrix},  & \delta d^{(I)}_{i}=\delta d^{(1)}_{N+1}
\\[10pt]
0, & \delta d^{(I)}_{i}=\delta d^{(2)}_{N+1}
\end{cases}.
\label{eq:Phi(s)xy Definition} 
\end{IEEEeqnarray}
\end{widetext}
where the $n_{s}$ is the refractive index of substrate(usually made by $\rm{SiO_{2}}$). Note that $\Delta n_{1}$ and $\Delta n_{2}$ in Eqs.(\ref{eq:Phi xx})-(\ref{eq:Phi(s)xy Definition}) is neglected since $\Delta n_{I}/n_x^I \ll 1$.

\subsubsection{The reflected light from the coating layers}

The entire coating layers can be described as a $4\times4$ matrix $\mathbf{M}$, connecting the optical field on the right of the coating to those on the left:
\begin{IEEEeqnarray}{rCl}
\begin{bmatrix}
 E_{in}\\E_{rx}\\0\\E_{ry}
\end{bmatrix}=\begin{bmatrix}
M_{11}&M_{12}&M_{13}&M_{14}
 \\M_{21}&M_{22}&M_{23}&M_{24}
 \\M_{31}&M_{32}&M_{33}&M_{34}
 \\M_{41}&M_{42}&M_{43}&M_{44}
\end{bmatrix}
\begin{bmatrix}
E_{tx}\\0\\E_{ty}\\0
\end{bmatrix}
\label{eq:Mmatrix},
\end{IEEEeqnarray}
where $E_{in}$ is incident light polarized along x-axis, $E_{rx},E_{ry}$ is the light field reflected by the coating layers, $E_{tx},E_{ty}$ is the transmitted light. Since we are interested in the variation of the polarization state, the only term that needs to be calculated is $E_{ry}$:
\begin{IEEEeqnarray}{rCl}
r_{y}&=&\frac{E_{ry}}{E_{in}}=\frac{M_{41}M_{33}-M_{43}M_{31}}{M_{11}M_{33}-M_{13}M_{31}},
\label{eq:ry}
\end{IEEEeqnarray}
which is obtained by solving Eq.\eqref{eq:Mmatrix}. 
The real part of $r_{y}$ represents the change in polarization angle, and the imaginary part represents the change in ellipticity (i.e., birefringence). It turns out that most of the $r_{y}$ is the imaginary part (see Section \ref{sec:sec3}) so that the real part can be neglected. The transfer matrix $\mathbf{M}$ describes the response of the light field with respect to the coating surface, therefore $\mathbf{M}$ depends on the rotation angle of the optical axis $\delta \theta$. Futhermore, the $\delta\theta$ is related to $\delta d$ by Eq.(\ref{eq:polarizaton angle change}), which leads to a relationship between the ellipticity fluctuations spectrum $S_{\psi}$ and the deformation spectrum of a single coating layer $S_{d_{i}^{(I)}}$:
\begin{IEEEeqnarray}{rCl}
S_{\psi}^{1/2}=\sqrt{\sum_{I=1}^{2}\sum_{i=1}^{N}\left(\frac{\partial r_{y}}{\partial d_{i}^{(I)}}\right)^{2} S_{d_{i}^{(I)}} } .
\label{eq:delata phi}
\end{IEEEeqnarray}
The fluctuation of the coating layer thickness $\delta d_{i}^{(I)}$ is typically a Brownian type thereby showing a Brownian spectrum of $S_{\psi}$.
Note that the summation is over the influence of different layers. We do not have the cross term like $S_{d_{i}^{(I)} d_{j}^{(J)}}$ is because that the deformations caused by Brownian motion in the different layers are independent of each other\cite{hong2013brownian}. 

\section{\label{sec:sec3}Calculated noise level}

\renewcommand{\arraystretch}{1.5}
\begin{table}[t]
\captionsetup{font={footnotesize},justification=RaggedRight}
\caption{\label{tab:table1}Typical parameters used in this paper.}
\begin{ruledtabular}
\begin{tabular}{p{4cm}ccc}
Parameters& Symbol & $\rm{GaAs}$&$\rm{AlGaAs}$ \\
\colrule

Refractive index & $n$ & 3.37\cite{adachi1993properties} & 2.90\cite{afromowitz1974refractive} \\
Poisson's ratio & $\sigma$ & \multicolumn{2}{c}{0.32\cite{adachi1993properties}}  \\
Young's modulus(GPa) & $Y$ & \multicolumn{2}{c}{100\cite{cole2013tenfold}}\\
Loss angle & $\phi$ & \multicolumn{2}{c}{$2.41 \times 10^{-5}$\cite{cole2011phonon}}\\
Temperature(K) & $T$ & \multicolumn{2}{c}{293}\\
Beam radius($\mu m$) & $r_{0}$ & \multicolumn{2}{c}{1500}\\ 
Photoelastic coefficient & $p_{13}$ & \multicolumn{2}{c}{-0.14\cite{dixon1967photoelastic}}\\ 
Photoelastic coefficient & $p_{63}$ & \multicolumn{2}{c}{-0.0014}\\ 
\end{tabular}
\end{ruledtabular}
\end{table}

In this section we present the noise level of the non-diagonal anisotropic Brownian photoelastic (NABP) effect using the parameters listed in Table\,\ref{tab:table1}, based on the theory established in Section \ref{sec:sec2}. As mentioned in Section \ref{sec:sec2A}, the $p_{63}$ in the crystalline coating materials may arise from strain induced during the deposition process, and there is currently no experimental measurement of $p_{63}$ in the coating material. On the other hand, 
the theoretical quantitative estimation of $p_{63}$ requires a first-principle density-functional calculation\,\cite{detraux2001photoelasticity,donadio2003photoelasticity}. In this paper, we focus on demonstrating the influence of the non-diagonal anisotropic Brownian photoelastic effect, and the value of $p_{63}=-0.0014$ is assumed to be one percent of $p_{13}$ and the values of $p_{63}$ in both coating materials are assumed to be the same. The real value of $p_{63}$ could be smaller than this assumption, which needs to be determined by further experimental test or more detailed density-functional calculation by the material science theorists.

\begin{figure}[t]
\includegraphics[width=8.6cm]{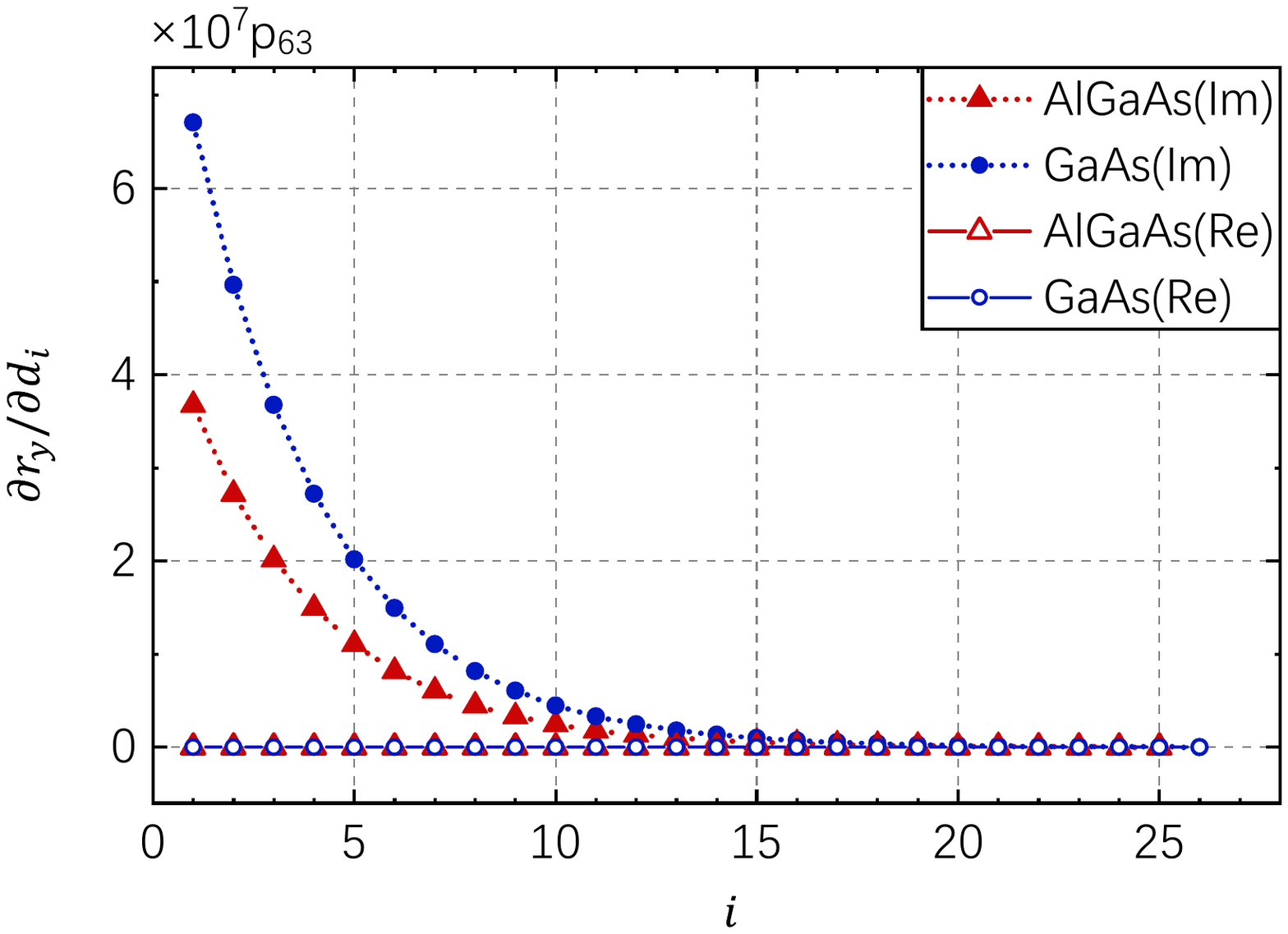}
\captionsetup{font={footnotesize}, justification=RaggedRight}
\caption{Influence of different layer thicknesses on the ellipticity for a 19 pairs of $\rm{GaAs}$/$\rm{AlGaAs}$ coated mirror. As previously mentioned, the imaginary part of $\partial r_{y}/\partial d_{i}$ represents the change in ellipticity of the light reflected from the mirror and the real part represents the change in polarization angle. The real part of $\partial r_{y}/\partial d_{i}$  is four orders of magnitude smaller than the imaginary part thereby can be ignored.}
\label{fig:Influence of different layer}
\end{figure} 

One can derive a complete expression for the $\mathbf{M}$-matrix by substituting Eqs.(\ref{eq:Phi_i}-\ref{eq:Phi(s)xy Definition}) into Eq.(\ref{eq:M matrix}). The contribution of the different layers to the ellipticity ($\partial r_{y} /\partial d_{i}$) is then obtained by Eq.(\ref{eq:ry}), as shown in Fig.\ref{fig:Influence of different layer}. We can see that the effect of the layer on the ellipticity decreases as the number of layers increases, so that only the first few layers contribute most to the variation of the ellipticity.

For a single coating layer, the thickness fluctuation due to the Brownian motion is obtained using Levin's approach\,\cite{levin1998internal,hong2013brownian}:
\begin{IEEEeqnarray}{rCl}
S_{d}(f)=\frac{8k_{B}T d_{i} \phi_{i} (1-\sigma_{i}-2\sigma_{i}^{2}) }{3\pi^{2} f Y_{i} (1-\sigma_{i})^{2} r_{0}^{2}},
\label{eq:d noise level}
\end{IEEEeqnarray}
where $k_{B}$ is the Boltzmann constant, $d_{i}$ is the thickness of this layer, $\phi_{i}$ is the loss angle, $\sigma_{i}$ is Poisson's ratio, $Y_{i}$ is Young's modulus, $r_{0}$ is the radius of the laser beam. 

\begin{figure}[t]
\includegraphics[width=8.6cm]{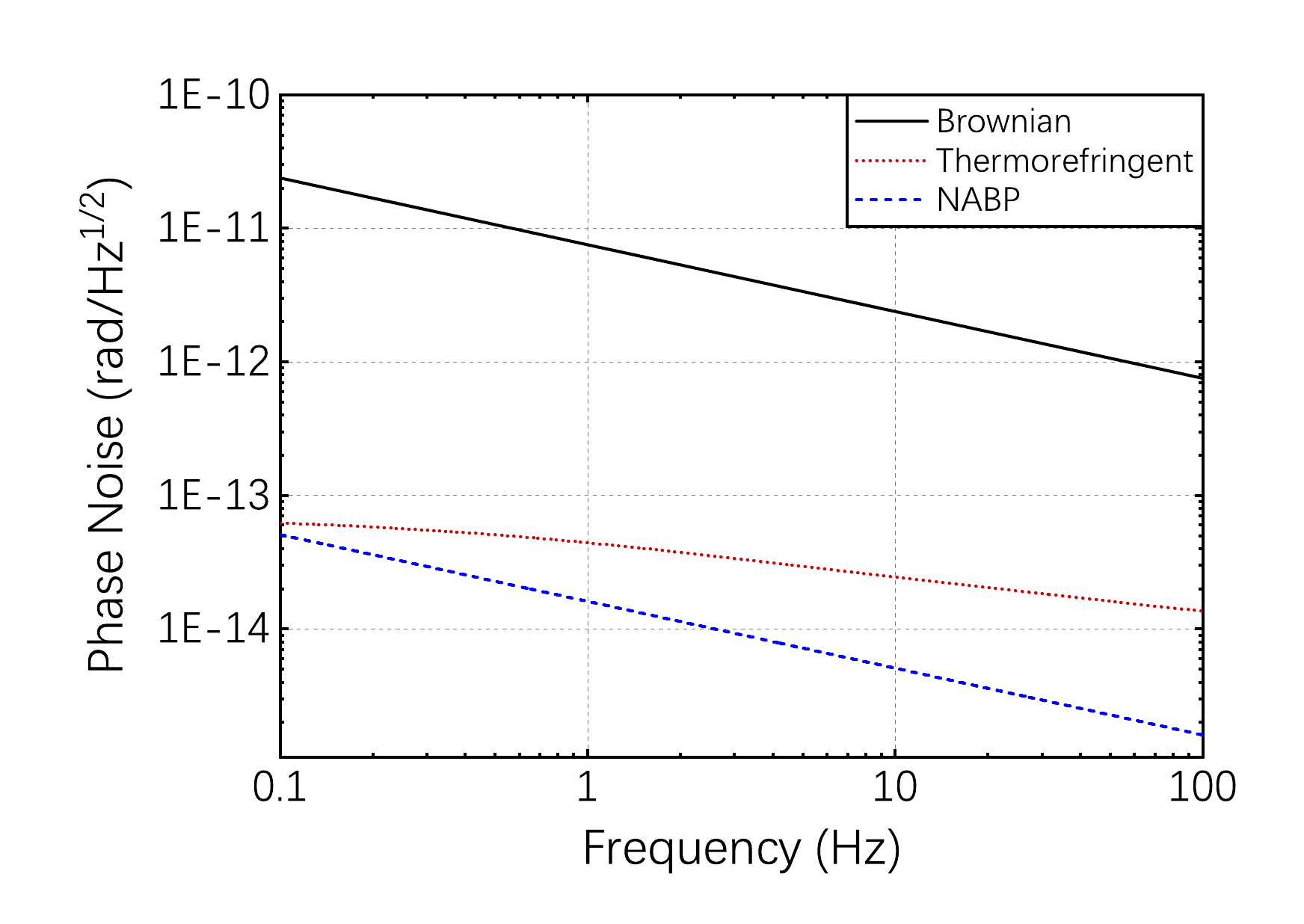}
\captionsetup{font={footnotesize}, justification=RaggedRight}
\caption{Non-diagonal anisotropic Brownian photoelastic noise (NABP) calculated for an 25-pair $\rm{GaAs}$/$\rm{AlGaAs}$ coated mirror. The comparison with the Brownian noise and thermorefringent noise is also shown.}
\label{fig:Noise level}
\end{figure}

The noise level of the NABP effect is obtained by substituting Eq.(\ref{eq:d noise level}) into Eq.(\ref{eq:delata phi}), as shown in Fig.\ref{fig:Noise level}. This result demonstrates that, using the sampled parameters for the $p_{63}$ in Table\,\ref{tab:table1}, the NABP noise is 
\begin{equation}
% \sqrt{S_{\psi\psi}}
S_{\psi(NABP)}^{1/2}=1.2\times 10^{-11} p_{63}f^{-1/2}/\sqrt{\text{Hz}},
\end{equation}
% Our experiment reported that the noise floor is limited by the intrinsic noise of the cavity mirror, and the exact mechanism of this noise is not yet clear. 
which is shown in Fig.\ref{fig:Noise level}, and the Brownian noise and the thermorefringent noise are also shown for comparison. Here the thermorefringent noise is \cite{kryhin2023thermorefringent}:
\begin{IEEEeqnarray}{rCl}
S_{\psi (TF)}=\left | r_{y}\frac{\pi}{2} \frac{n_{2}\varepsilon'^{(1)}_{xy}+n_{1}\varepsilon'^{(2)}_{xy}}{n_{1}n_{2}(n_{1}^{2}-n_{2}^{2})} \right|^{2} S_{uu}(f),
\label{eq:thermorefringent noise}
\end{IEEEeqnarray}
where $\varepsilon'_{xy}=10 \rm{ppm/K} $ is thermorefractive coeffcient, $r_{y}$ is reflection amplitude for $y$-polarization and $S_{uu}$ is the power spectral density of temperature averaged over the beam spot:
\begin{IEEEeqnarray}{rCl}
S_{uu}(f)=\frac{2k_{B}T^{2}}{\pi r_{0}c_{V}D} \mathrm{Re}\left \{ \int\limits_{0}^{\infty } du\frac{ue^{-u^{2}/2}}{\sqrt{u^{2}-ir_{0}^{2}f/D} } \right \},
\end{IEEEeqnarray}
where $c_{V}=1.6\times 10^{6} \rm{J(mK)^{-1}}$ is heat capacity per unit volume, $\kappa=1.38\rm{W(mK)^{-1}}$ is the thermal conductivity and $D=\kappa /c_{V}$ is thermal diffusivity.

\section{\label{sec:sec4}Conclusions}

For experiments such as the measurement of vacuum magnetic birefringence, the measurement principle is to amplify the signal through a resonant cavity, which is then detected by a polarimetry. This polarimetry detects the phase difference between the two polarization directions, so that the phase noise is cancelled out. Thus, the sensitivity of these experiments is not limited by the commonly discussed thermal phase noise\,(Brownian, thermo-optic noise) but by the thermal birefringent noise. Experimental measurements and theoretical investigations of the anisotropy of these coating materials are important. 

In this work, we present a comprehensive analysis of a noval kind of birefringent noise induced by the Brownian motion of the mirror coating. In particular, we discussed the possible 
influence of a novel noise source originating from the non-diagonal anisotropic photoelastic effect, which may limited the sensitivity of precision polarimetry experiments. The strain deformation caused by Brownian motion in a single coating layer induces a rotation of the intrinsic birefringence due to the non-diagonal anisotropic photoelastic effect. By extending these findings to multilayer surfaces using the matrix method, we elucidate the corresponding fluctuations in the light reflected from the mirror.

For the $\rm{GaAs}$-$\rm{AlGaAs}$ coating, the noise level caused by the non-diagonal anisotropic Brownian photoelastic\,(NABP) effect is calculated to be $1.2\times 10^{-11} p_{63} f^{-1/2}/\rm{Hz}^{1/2} $ using the sampled parameters in Table\,\ref{tab:table1}. The non-diagonal anisotropic Brownian photoelastic noise level is decisively dependent on the value of $p_{63}$. Assuming that $p_{63}$ is one percent of the $p_{13}$, the NABP noise level is of the same order of magnitude as the thermorefringent noise. In addition, the anisotropy in the different parts of the coating can introduce loss in reflectivity\,\cite{vyatchanin2020loss}. This will generate noise according to the fluctuation-dissipation theorem, which could be a stronger effect than the NABP noise effect. This anisotropy can be evaluated through measuring the intrinsic biefringence of different parts of the coating surface, hence further compensated probably. 
The source of the $p_{63}$ and other anisotropies of the coating materials could be the strain induced during the manufacturing process, the specific mechanisms and the evaluation of $p_{63}$ for the coating material require further experimental and theoretical investigation. \\

\begin{acknowledgments}
The authors thank Professor Chunnong Zhao, Dr. Haoyu Wang, Dr. Serhii Kryhin and Dr. Evan D. Hall for very helpful discussions. They also want to thank the anonymous referees for giving very useful suggestions to improve our manuscropt. In addition, the authors also thank Dr. Lin Zhu and other members of the MOE Key Laboratory of Fundamental Physical Quantities Measurement for useful discussions. This work is supported by the the National Key R\&D Program of China\,(2023YFC2205801), National Natural Science Foundation of China under Grant Nos.12150012,12175317,11925503.
\end{acknowledgments}

\bibliography{apssamp}

\begin{thebibliography}{36}%
\makeatletter
\providecommand \@ifxundefined [1]{%
 \@ifx{#1\undefined}
}%
\providecommand \@ifnum [1]{%
 \ifnum #1\expandafter \@firstoftwo
 \else \expandafter \@secondoftwo
 \fi
}%
\providecommand \@ifx [1]{%
 \ifx #1\expandafter \@firstoftwo
 \else \expandafter \@secondoftwo
 \fi
}%
\providecommand \natexlab [1]{#1}%
\providecommand \enquote  [1]{``#1''}%
\providecommand \bibnamefont  [1]{#1}%
\providecommand \bibfnamefont [1]{#1}%
\providecommand \citenamefont [1]{#1}%
\providecommand \href@noop [0]{\@secondoftwo}%
\providecommand \href [0]{\begingroup \@sanitize@url \@href}%
\providecommand \@href[1]{\@@startlink{#1}\@@href}%
\providecommand \@@href[1]{\endgroup#1\@@endlink}%
\providecommand \@sanitize@url [0]{\catcode `\\12\catcode `\$12\catcode
  `\&12\catcode `\#12\catcode `\^12\catcode `\_12\catcode `\%12\relax}%
\providecommand \@@startlink[1]{}%
\providecommand \@@endlink[0]{}%
\providecommand \url  [0]{\begingroup\@sanitize@url \@url }%
\providecommand \@url [1]{\endgroup\@href {#1}{\urlprefix }}%
\providecommand \urlprefix  [0]{URL }%
\providecommand \Eprint [0]{\href }%
\providecommand \doibase [0]{https://doi.org/}%
\providecommand \selectlanguage [0]{\@gobble}%
\providecommand \bibinfo  [0]{\@secondoftwo}%
\providecommand \bibfield  [0]{\@secondoftwo}%
\providecommand \translation [1]{[#1]}%
\providecommand \BibitemOpen [0]{}%
\providecommand \bibitemStop [0]{}%
\providecommand \bibitemNoStop [0]{.\EOS\space}%
\providecommand \EOS [0]{\spacefactor3000\relax}%
\providecommand \BibitemShut  [1]{\csname bibitem#1\endcsname}%
\let\auto@bib@innerbib\@empty
%</preamble>
\bibitem [{\citenamefont {Harry}\ \emph {et~al.}(2002)\citenamefont {Harry},
  \citenamefont {Gretarsson}, \citenamefont {Saulson}, \citenamefont
  {Kittelberger}, \citenamefont {Penn}, \citenamefont {Startin}, \citenamefont
  {Rowan}, \citenamefont {Fejer}, \citenamefont {Crooks}, \citenamefont
  {Cagnoli} \emph {et~al.}}]{harry2002thermal}%
  \BibitemOpen
  \bibfield  {author} {\bibinfo {author} {\bibfnamefont {G.~M.}\ \bibnamefont
  {Harry}}, \bibinfo {author} {\bibfnamefont {A.~M.}\ \bibnamefont
  {Gretarsson}}, \bibinfo {author} {\bibfnamefont {P.~R.}\ \bibnamefont
  {Saulson}}, \bibinfo {author} {\bibfnamefont {S.~E.}\ \bibnamefont
  {Kittelberger}}, \bibinfo {author} {\bibfnamefont {S.~D.}\ \bibnamefont
  {Penn}}, \bibinfo {author} {\bibfnamefont {W.~J.}\ \bibnamefont {Startin}},
  \bibinfo {author} {\bibfnamefont {S.}~\bibnamefont {Rowan}}, \bibinfo
  {author} {\bibfnamefont {M.~M.}\ \bibnamefont {Fejer}}, \bibinfo {author}
  {\bibfnamefont {D.}~\bibnamefont {Crooks}}, \bibinfo {author} {\bibfnamefont
  {G.}~\bibnamefont {Cagnoli}}, \emph {et~al.},\ }\bibfield  {title} {\bibinfo
  {title} {Thermal noise in interferometric gravitational wave detectors due to
  dielectric optical coatings},\ }\href@noop {} {\bibfield  {journal} {\bibinfo
   {journal} {Classical and Quantum Gravity}\ }\textbf {\bibinfo {volume}
  {19}},\ \bibinfo {pages} {897} (\bibinfo {year} {2002})}\BibitemShut
  {NoStop}%
\bibitem [{\citenamefont {Villar}\ \emph {et~al.}(2010)\citenamefont {Villar},
  \citenamefont {Black}, \citenamefont {DeSalvo}, \citenamefont {Libbrecht},
  \citenamefont {Michel}, \citenamefont {Morgado}, \citenamefont {Pinard},
  \citenamefont {Pinto}, \citenamefont {Pierro}, \citenamefont {Galdi} \emph
  {et~al.}}]{villar2010measurement}%
  \BibitemOpen
  \bibfield  {author} {\bibinfo {author} {\bibfnamefont {A.~E.}\ \bibnamefont
  {Villar}}, \bibinfo {author} {\bibfnamefont {E.~D.}\ \bibnamefont {Black}},
  \bibinfo {author} {\bibfnamefont {R.}~\bibnamefont {DeSalvo}}, \bibinfo
  {author} {\bibfnamefont {K.~G.}\ \bibnamefont {Libbrecht}}, \bibinfo {author}
  {\bibfnamefont {C.}~\bibnamefont {Michel}}, \bibinfo {author} {\bibfnamefont
  {N.}~\bibnamefont {Morgado}}, \bibinfo {author} {\bibfnamefont
  {L.}~\bibnamefont {Pinard}}, \bibinfo {author} {\bibfnamefont {I.~M.}\
  \bibnamefont {Pinto}}, \bibinfo {author} {\bibfnamefont {V.}~\bibnamefont
  {Pierro}}, \bibinfo {author} {\bibfnamefont {V.}~\bibnamefont {Galdi}}, \emph
  {et~al.},\ }\bibfield  {title} {\bibinfo {title} {Measurement of thermal
  noise in multilayer coatings with optimized layer thickness},\ }\href@noop {}
  {\bibfield  {journal} {\bibinfo  {journal} {Physical Review D}\ }\textbf
  {\bibinfo {volume} {81}},\ \bibinfo {pages} {122001} (\bibinfo {year}
  {2010})}\BibitemShut {NoStop}%
\bibitem [{\citenamefont {Chalermsongsak}\ \emph {et~al.}(2014)\citenamefont
  {Chalermsongsak}, \citenamefont {Seifert}, \citenamefont {Hall},
  \citenamefont {Arai}, \citenamefont {Gustafson},\ and\ \citenamefont
  {Adhikari}}]{chalermsongsak2014broadband}%
  \BibitemOpen
  \bibfield  {author} {\bibinfo {author} {\bibfnamefont {T.}~\bibnamefont
  {Chalermsongsak}}, \bibinfo {author} {\bibfnamefont {F.}~\bibnamefont
  {Seifert}}, \bibinfo {author} {\bibfnamefont {E.~D.}\ \bibnamefont {Hall}},
  \bibinfo {author} {\bibfnamefont {K.}~\bibnamefont {Arai}}, \bibinfo {author}
  {\bibfnamefont {E.~K.}\ \bibnamefont {Gustafson}},\ and\ \bibinfo {author}
  {\bibfnamefont {R.~X.}\ \bibnamefont {Adhikari}},\ }\bibfield  {title}
  {\bibinfo {title} {Broadband measurement of coating thermal noise in rigid
  fabry--p{\'e}rot cavities},\ }\href@noop {} {\bibfield  {journal} {\bibinfo
  {journal} {Metrologia}\ }\textbf {\bibinfo {volume} {52}},\ \bibinfo {pages}
  {17} (\bibinfo {year} {2014})}\BibitemShut {NoStop}%
\bibitem [{\citenamefont {Gras}\ \emph {et~al.}(2017)\citenamefont {Gras},
  \citenamefont {Yu}, \citenamefont {Yam}, \citenamefont {Martynov},\ and\
  \citenamefont {Evans}}]{gras2017audio}%
  \BibitemOpen
  \bibfield  {author} {\bibinfo {author} {\bibfnamefont {S.}~\bibnamefont
  {Gras}}, \bibinfo {author} {\bibfnamefont {H.}~\bibnamefont {Yu}}, \bibinfo
  {author} {\bibfnamefont {W.}~\bibnamefont {Yam}}, \bibinfo {author}
  {\bibfnamefont {D.}~\bibnamefont {Martynov}},\ and\ \bibinfo {author}
  {\bibfnamefont {M.}~\bibnamefont {Evans}},\ }\bibfield  {title} {\bibinfo
  {title} {Audio-band coating thermal noise measurement for advanced ligo with
  a multimode optical resonator},\ }\href@noop {} {\bibfield  {journal}
  {\bibinfo  {journal} {Physical Review D}\ }\textbf {\bibinfo {volume} {95}},\
  \bibinfo {pages} {022001} (\bibinfo {year} {2017})}\BibitemShut {NoStop}%
\bibitem [{\citenamefont {Granata}\ \emph {et~al.}(2020)\citenamefont
  {Granata}, \citenamefont {Amato}, \citenamefont {Cagnoli}, \citenamefont
  {Coulon}, \citenamefont {Degallaix}, \citenamefont {Forest}, \citenamefont
  {Mereni}, \citenamefont {Michel}, \citenamefont {Pinard}, \citenamefont
  {Sassolas} \emph {et~al.}}]{granata2020progress}%
  \BibitemOpen
  \bibfield  {author} {\bibinfo {author} {\bibfnamefont {M.}~\bibnamefont
  {Granata}}, \bibinfo {author} {\bibfnamefont {A.}~\bibnamefont {Amato}},
  \bibinfo {author} {\bibfnamefont {G.}~\bibnamefont {Cagnoli}}, \bibinfo
  {author} {\bibfnamefont {M.}~\bibnamefont {Coulon}}, \bibinfo {author}
  {\bibfnamefont {J.}~\bibnamefont {Degallaix}}, \bibinfo {author}
  {\bibfnamefont {D.}~\bibnamefont {Forest}}, \bibinfo {author} {\bibfnamefont
  {L.}~\bibnamefont {Mereni}}, \bibinfo {author} {\bibfnamefont
  {C.}~\bibnamefont {Michel}}, \bibinfo {author} {\bibfnamefont
  {L.}~\bibnamefont {Pinard}}, \bibinfo {author} {\bibfnamefont
  {B.}~\bibnamefont {Sassolas}}, \emph {et~al.},\ }\bibfield  {title} {\bibinfo
  {title} {Progress in the measurement and reduction of thermal noise in
  optical coatings for gravitational-wave detectors},\ }\href@noop {}
  {\bibfield  {journal} {\bibinfo  {journal} {Applied optics}\ }\textbf
  {\bibinfo {volume} {59}},\ \bibinfo {pages} {A229} (\bibinfo {year}
  {2020})}\BibitemShut {NoStop}%
\bibitem [{\citenamefont {Numata}\ \emph {et~al.}(2003)\citenamefont {Numata},
  \citenamefont {Ando}, \citenamefont {Yamamoto}, \citenamefont {Otsuka},\ and\
  \citenamefont {Tsubono}}]{numata2003wide}%
  \BibitemOpen
  \bibfield  {author} {\bibinfo {author} {\bibfnamefont {K.}~\bibnamefont
  {Numata}}, \bibinfo {author} {\bibfnamefont {M.}~\bibnamefont {Ando}},
  \bibinfo {author} {\bibfnamefont {K.}~\bibnamefont {Yamamoto}}, \bibinfo
  {author} {\bibfnamefont {S.}~\bibnamefont {Otsuka}},\ and\ \bibinfo {author}
  {\bibfnamefont {K.}~\bibnamefont {Tsubono}},\ }\bibfield  {title} {\bibinfo
  {title} {Wide-band direct measurement of thermal fluctuations in an
  interferometer},\ }\href@noop {} {\bibfield  {journal} {\bibinfo  {journal}
  {Physical review letters}\ }\textbf {\bibinfo {volume} {91}},\ \bibinfo
  {pages} {260602} (\bibinfo {year} {2003})}\BibitemShut {NoStop}%
\bibitem [{\citenamefont {Numata}\ \emph {et~al.}(2004)\citenamefont {Numata},
  \citenamefont {Kemery},\ and\ \citenamefont {Camp}}]{numata2004thermal}%
  \BibitemOpen
  \bibfield  {author} {\bibinfo {author} {\bibfnamefont {K.}~\bibnamefont
  {Numata}}, \bibinfo {author} {\bibfnamefont {A.}~\bibnamefont {Kemery}},\
  and\ \bibinfo {author} {\bibfnamefont {J.}~\bibnamefont {Camp}},\ }\bibfield
  {title} {\bibinfo {title} {Thermal-noise limit in the frequency stabilization
  of lasers with rigid cavities},\ }\href@noop {} {\bibfield  {journal}
  {\bibinfo  {journal} {Physical review letters}\ }\textbf {\bibinfo {volume}
  {93}},\ \bibinfo {pages} {250602} (\bibinfo {year} {2004})}\BibitemShut
  {NoStop}%
\bibitem [{\citenamefont {Notcutt}\ \emph {et~al.}(2006)\citenamefont
  {Notcutt}, \citenamefont {Ma}, \citenamefont {Ludlow}, \citenamefont
  {Foreman}, \citenamefont {Ye},\ and\ \citenamefont
  {Hall}}]{notcutt2006contribution}%
  \BibitemOpen
  \bibfield  {author} {\bibinfo {author} {\bibfnamefont {M.}~\bibnamefont
  {Notcutt}}, \bibinfo {author} {\bibfnamefont {L.-S.}\ \bibnamefont {Ma}},
  \bibinfo {author} {\bibfnamefont {A.~D.}\ \bibnamefont {Ludlow}}, \bibinfo
  {author} {\bibfnamefont {S.~M.}\ \bibnamefont {Foreman}}, \bibinfo {author}
  {\bibfnamefont {J.}~\bibnamefont {Ye}},\ and\ \bibinfo {author}
  {\bibfnamefont {J.~L.}\ \bibnamefont {Hall}},\ }\bibfield  {title} {\bibinfo
  {title} {Contribution of thermal noise to frequency stability of rigid
  optical cavity via hertz-linewidth lasers},\ }\href@noop {} {\bibfield
  {journal} {\bibinfo  {journal} {Physical Review A}\ }\textbf {\bibinfo
  {volume} {73}},\ \bibinfo {pages} {031804} (\bibinfo {year}
  {2006})}\BibitemShut {NoStop}%
\bibitem [{\citenamefont {Matei}\ \emph {et~al.}(2017)\citenamefont {Matei},
  \citenamefont {Legero}, \citenamefont {H{\"a}fner}, \citenamefont {Grebing},
  \citenamefont {Weyrich}, \citenamefont {Zhang}, \citenamefont {Sonderhouse},
  \citenamefont {Robinson}, \citenamefont {Ye}, \citenamefont {Riehle} \emph
  {et~al.}}]{matei20171}%
  \BibitemOpen
  \bibfield  {author} {\bibinfo {author} {\bibfnamefont {D.}~\bibnamefont
  {Matei}}, \bibinfo {author} {\bibfnamefont {T.}~\bibnamefont {Legero}},
  \bibinfo {author} {\bibfnamefont {S.}~\bibnamefont {H{\"a}fner}}, \bibinfo
  {author} {\bibfnamefont {C.}~\bibnamefont {Grebing}}, \bibinfo {author}
  {\bibfnamefont {R.}~\bibnamefont {Weyrich}}, \bibinfo {author} {\bibfnamefont
  {W.}~\bibnamefont {Zhang}}, \bibinfo {author} {\bibfnamefont
  {L.}~\bibnamefont {Sonderhouse}}, \bibinfo {author} {\bibfnamefont
  {J.}~\bibnamefont {Robinson}}, \bibinfo {author} {\bibfnamefont
  {J.}~\bibnamefont {Ye}}, \bibinfo {author} {\bibfnamefont {F.}~\bibnamefont
  {Riehle}}, \emph {et~al.},\ }\bibfield  {title} {\bibinfo {title} {1.5 $\mu$
  m lasers with sub-10 mhz linewidth},\ }\href@noop {} {\bibfield  {journal}
  {\bibinfo  {journal} {Physical review letters}\ }\textbf {\bibinfo {volume}
  {118}},\ \bibinfo {pages} {263202} (\bibinfo {year} {2017})}\BibitemShut
  {NoStop}%
\bibitem [{\citenamefont {Ma}\ \emph {et~al.}(2020)\citenamefont {Ma},
  \citenamefont {Liu}, \citenamefont {Wei}, \citenamefont {Yuan}, \citenamefont
  {Hao}, \citenamefont {Deng}, \citenamefont {Che}, \citenamefont {Xu},
  \citenamefont {Cheng}, \citenamefont {Wang} \emph
  {et~al.}}]{ma2020investigation}%
  \BibitemOpen
  \bibfield  {author} {\bibinfo {author} {\bibfnamefont {Z.}~\bibnamefont
  {Ma}}, \bibinfo {author} {\bibfnamefont {H.}~\bibnamefont {Liu}}, \bibinfo
  {author} {\bibfnamefont {W.}~\bibnamefont {Wei}}, \bibinfo {author}
  {\bibfnamefont {W.}~\bibnamefont {Yuan}}, \bibinfo {author} {\bibfnamefont
  {P.}~\bibnamefont {Hao}}, \bibinfo {author} {\bibfnamefont {Z.}~\bibnamefont
  {Deng}}, \bibinfo {author} {\bibfnamefont {H.}~\bibnamefont {Che}}, \bibinfo
  {author} {\bibfnamefont {Z.}~\bibnamefont {Xu}}, \bibinfo {author}
  {\bibfnamefont {F.}~\bibnamefont {Cheng}}, \bibinfo {author} {\bibfnamefont
  {Z.}~\bibnamefont {Wang}}, \emph {et~al.},\ }\bibfield  {title} {\bibinfo
  {title} {Investigation of experimental issues concerning successful operation
  of quantum-logic-based 27 al+ ion optical clock},\ }\href@noop {} {\bibfield
  {journal} {\bibinfo  {journal} {Applied Physics B}\ }\textbf {\bibinfo
  {volume} {126}},\ \bibinfo {pages} {129} (\bibinfo {year}
  {2020})}\BibitemShut {NoStop}%
\bibitem [{\citenamefont {Yu}\ \emph {et~al.}(2023)\citenamefont {Yu},
  \citenamefont {H{\"a}fner}, \citenamefont {Legero}, \citenamefont {Herbers},
  \citenamefont {Nicolodi}, \citenamefont {Ma}, \citenamefont {Riehle},
  \citenamefont {Sterr}, \citenamefont {Kedar}, \citenamefont {Robinson} \emph
  {et~al.}}]{yu2023excess}%
  \BibitemOpen
  \bibfield  {author} {\bibinfo {author} {\bibfnamefont {J.}~\bibnamefont
  {Yu}}, \bibinfo {author} {\bibfnamefont {S.}~\bibnamefont {H{\"a}fner}},
  \bibinfo {author} {\bibfnamefont {T.}~\bibnamefont {Legero}}, \bibinfo
  {author} {\bibfnamefont {S.}~\bibnamefont {Herbers}}, \bibinfo {author}
  {\bibfnamefont {D.}~\bibnamefont {Nicolodi}}, \bibinfo {author}
  {\bibfnamefont {C.~Y.}\ \bibnamefont {Ma}}, \bibinfo {author} {\bibfnamefont
  {F.}~\bibnamefont {Riehle}}, \bibinfo {author} {\bibfnamefont
  {U.}~\bibnamefont {Sterr}}, \bibinfo {author} {\bibfnamefont
  {D.}~\bibnamefont {Kedar}}, \bibinfo {author} {\bibfnamefont {J.~M.}\
  \bibnamefont {Robinson}}, \emph {et~al.},\ }\bibfield  {title} {\bibinfo
  {title} {Excess noise and photoinduced effects in highly reflective
  crystalline mirror coatings},\ }\href@noop {} {\bibfield  {journal} {\bibinfo
   {journal} {Physical Review X}\ }\textbf {\bibinfo {volume} {13}},\ \bibinfo
  {pages} {041002} (\bibinfo {year} {2023})}\BibitemShut {NoStop}%
\bibitem [{\citenamefont {Levin}(1998)}]{levin1998internal}%
  \BibitemOpen
  \bibfield  {author} {\bibinfo {author} {\bibfnamefont {Y.}~\bibnamefont
  {Levin}},\ }\bibfield  {title} {\bibinfo {title} {Internal thermal noise in
  the ligo test masses: A direct approach},\ }\href@noop {} {\bibfield
  {journal} {\bibinfo  {journal} {Physical Review D}\ }\textbf {\bibinfo
  {volume} {57}},\ \bibinfo {pages} {659} (\bibinfo {year} {1998})}\BibitemShut
  {NoStop}%
\bibitem [{\citenamefont {Evans}\ \emph {et~al.}(2008)\citenamefont {Evans},
  \citenamefont {Ballmer}, \citenamefont {Fejer}, \citenamefont {Fritschel},
  \citenamefont {Harry},\ and\ \citenamefont {Ogin}}]{evans2008thermo}%
  \BibitemOpen
  \bibfield  {author} {\bibinfo {author} {\bibfnamefont {M.}~\bibnamefont
  {Evans}}, \bibinfo {author} {\bibfnamefont {S.}~\bibnamefont {Ballmer}},
  \bibinfo {author} {\bibfnamefont {M.}~\bibnamefont {Fejer}}, \bibinfo
  {author} {\bibfnamefont {P.}~\bibnamefont {Fritschel}}, \bibinfo {author}
  {\bibfnamefont {G.}~\bibnamefont {Harry}},\ and\ \bibinfo {author}
  {\bibfnamefont {G.}~\bibnamefont {Ogin}},\ }\bibfield  {title} {\bibinfo
  {title} {Thermo-optic noise in coated mirrors for high-precision optical
  measurements},\ }\href@noop {} {\bibfield  {journal} {\bibinfo  {journal}
  {Physical Review D}\ }\textbf {\bibinfo {volume} {78}},\ \bibinfo {pages}
  {102003} (\bibinfo {year} {2008})}\BibitemShut {NoStop}%
\bibitem [{\citenamefont {Hong}\ \emph {et~al.}(2013)\citenamefont {Hong},
  \citenamefont {Yang}, \citenamefont {Gustafson}, \citenamefont {Adhikari},\
  and\ \citenamefont {Chen}}]{hong2013brownian}%
  \BibitemOpen
  \bibfield  {author} {\bibinfo {author} {\bibfnamefont {T.}~\bibnamefont
  {Hong}}, \bibinfo {author} {\bibfnamefont {H.}~\bibnamefont {Yang}}, \bibinfo
  {author} {\bibfnamefont {E.~K.}\ \bibnamefont {Gustafson}}, \bibinfo {author}
  {\bibfnamefont {R.~X.}\ \bibnamefont {Adhikari}},\ and\ \bibinfo {author}
  {\bibfnamefont {Y.}~\bibnamefont {Chen}},\ }\bibfield  {title} {\bibinfo
  {title} {Brownian thermal noise in multilayer coated mirrors},\ }\href@noop
  {} {\bibfield  {journal} {\bibinfo  {journal} {Physical Review D}\ }\textbf
  {\bibinfo {volume} {87}},\ \bibinfo {pages} {082001} (\bibinfo {year}
  {2013})}\BibitemShut {NoStop}%
\bibitem [{\citenamefont {Cameron}\ \emph {et~al.}(1993)\citenamefont
  {Cameron}, \citenamefont {Cantatore}, \citenamefont {Melissinos},
  \citenamefont {Ruoso}, \citenamefont {Semertzidis}, \citenamefont {Halama},
  \citenamefont {Lazarus}, \citenamefont {Prodell}, \citenamefont {Nezrick},
  \citenamefont {Rizzo} \emph {et~al.}}]{cameron1993search}%
  \BibitemOpen
  \bibfield  {author} {\bibinfo {author} {\bibfnamefont {R.}~\bibnamefont
  {Cameron}}, \bibinfo {author} {\bibfnamefont {G.}~\bibnamefont {Cantatore}},
  \bibinfo {author} {\bibfnamefont {A.}~\bibnamefont {Melissinos}}, \bibinfo
  {author} {\bibfnamefont {G.}~\bibnamefont {Ruoso}}, \bibinfo {author}
  {\bibfnamefont {Y.}~\bibnamefont {Semertzidis}}, \bibinfo {author}
  {\bibfnamefont {H.}~\bibnamefont {Halama}}, \bibinfo {author} {\bibfnamefont
  {D.}~\bibnamefont {Lazarus}}, \bibinfo {author} {\bibfnamefont
  {A.}~\bibnamefont {Prodell}}, \bibinfo {author} {\bibfnamefont
  {F.}~\bibnamefont {Nezrick}}, \bibinfo {author} {\bibfnamefont
  {C.}~\bibnamefont {Rizzo}}, \emph {et~al.},\ }\bibfield  {title} {\bibinfo
  {title} {Search for nearly massless, weakly coupled particles by optical
  techniques},\ }\href@noop {} {\bibfield  {journal} {\bibinfo  {journal}
  {Physical Review D}\ }\textbf {\bibinfo {volume} {47}},\ \bibinfo {pages}
  {3707} (\bibinfo {year} {1993})}\BibitemShut {NoStop}%
\bibitem [{\citenamefont {Ejlli}\ \emph {et~al.}(2020)\citenamefont {Ejlli},
  \citenamefont {Della~Valle}, \citenamefont {Gastaldi}, \citenamefont
  {Messineo}, \citenamefont {Pengo}, \citenamefont {Ruoso},\ and\ \citenamefont
  {Zavattini}}]{ejlli2020pvlas}%
  \BibitemOpen
  \bibfield  {author} {\bibinfo {author} {\bibfnamefont {A.}~\bibnamefont
  {Ejlli}}, \bibinfo {author} {\bibfnamefont {F.}~\bibnamefont {Della~Valle}},
  \bibinfo {author} {\bibfnamefont {U.}~\bibnamefont {Gastaldi}}, \bibinfo
  {author} {\bibfnamefont {G.}~\bibnamefont {Messineo}}, \bibinfo {author}
  {\bibfnamefont {R.}~\bibnamefont {Pengo}}, \bibinfo {author} {\bibfnamefont
  {G.}~\bibnamefont {Ruoso}},\ and\ \bibinfo {author} {\bibfnamefont
  {G.}~\bibnamefont {Zavattini}},\ }\bibfield  {title} {\bibinfo {title} {The
  pvlas experiment: A 25 year effort to measure vacuum magnetic
  birefringence},\ }\href@noop {} {\bibfield  {journal} {\bibinfo  {journal}
  {Physics Reports}\ }\textbf {\bibinfo {volume} {871}},\ \bibinfo {pages} {1}
  (\bibinfo {year} {2020})}\BibitemShut {NoStop}%
\bibitem [{\citenamefont {Fan}\ \emph {et~al.}(2017)\citenamefont {Fan},
  \citenamefont {Kamioka}, \citenamefont {Inada}, \citenamefont {Yamazaki},
  \citenamefont {Namba}, \citenamefont {Asai}, \citenamefont {Omachi},
  \citenamefont {Yoshioka}, \citenamefont {Kuwata-Gonokami}, \citenamefont
  {Matsuo} \emph {et~al.}}]{fan2017oval}%
  \BibitemOpen
  \bibfield  {author} {\bibinfo {author} {\bibfnamefont {X.}~\bibnamefont
  {Fan}}, \bibinfo {author} {\bibfnamefont {S.}~\bibnamefont {Kamioka}},
  \bibinfo {author} {\bibfnamefont {T.}~\bibnamefont {Inada}}, \bibinfo
  {author} {\bibfnamefont {T.}~\bibnamefont {Yamazaki}}, \bibinfo {author}
  {\bibfnamefont {T.}~\bibnamefont {Namba}}, \bibinfo {author} {\bibfnamefont
  {S.}~\bibnamefont {Asai}}, \bibinfo {author} {\bibfnamefont {J.}~\bibnamefont
  {Omachi}}, \bibinfo {author} {\bibfnamefont {K.}~\bibnamefont {Yoshioka}},
  \bibinfo {author} {\bibfnamefont {M.}~\bibnamefont {Kuwata-Gonokami}},
  \bibinfo {author} {\bibfnamefont {A.}~\bibnamefont {Matsuo}}, \emph
  {et~al.},\ }\bibfield  {title} {\bibinfo {title} {The oval experiment: a new
  experiment to measure vacuum magnetic birefringence using high repetition
  pulsed magnets},\ }\href@noop {} {\bibfield  {journal} {\bibinfo  {journal}
  {The European Physical Journal D}\ }\textbf {\bibinfo {volume} {71}},\
  \bibinfo {pages} {1} (\bibinfo {year} {2017})}\BibitemShut {NoStop}%
\bibitem [{\citenamefont {Agil}\ \emph {et~al.}(2022)\citenamefont {Agil},
  \citenamefont {Battesti},\ and\ \citenamefont {Rizzo}}]{agil2022vacuum}%
  \BibitemOpen
  \bibfield  {author} {\bibinfo {author} {\bibfnamefont {J.}~\bibnamefont
  {Agil}}, \bibinfo {author} {\bibfnamefont {R.}~\bibnamefont {Battesti}},\
  and\ \bibinfo {author} {\bibfnamefont {C.}~\bibnamefont {Rizzo}},\ }\bibfield
   {title} {\bibinfo {title} {Vacuum birefringence experiments: optical
  noise},\ }\href@noop {} {\bibfield  {journal} {\bibinfo  {journal} {The
  European Physical Journal D}\ }\textbf {\bibinfo {volume} {76}},\ \bibinfo
  {pages} {192} (\bibinfo {year} {2022})}\BibitemShut {NoStop}%
\bibitem [{\citenamefont {Chen}\ \emph {et~al.}(2007)\citenamefont {Chen},
  \citenamefont {Mei},\ and\ \citenamefont {Ni}}]{chen2007q}%
  \BibitemOpen
  \bibfield  {author} {\bibinfo {author} {\bibfnamefont {S.-J.}\ \bibnamefont
  {Chen}}, \bibinfo {author} {\bibfnamefont {H.-H.}\ \bibnamefont {Mei}},\ and\
  \bibinfo {author} {\bibfnamefont {W.-T.}\ \bibnamefont {Ni}},\ }\bibfield
  {title} {\bibinfo {title} {Q \& a experiment to search for vacuum dichroism,
  pseudoscalar--photon interaction and millicharged fermions},\ }\href@noop {}
  {\bibfield  {journal} {\bibinfo  {journal} {Modern Physics Letters A}\
  }\textbf {\bibinfo {volume} {22}},\ \bibinfo {pages} {2815} (\bibinfo {year}
  {2007})}\BibitemShut {NoStop}%
\bibitem [{\citenamefont {Ehret}\ \emph {et~al.}(2010)\citenamefont {Ehret},
  \citenamefont {Frede}, \citenamefont {Ghazaryan}, \citenamefont
  {Hildebrandt}, \citenamefont {Knabbe}, \citenamefont {Kracht}, \citenamefont
  {Lindner}, \citenamefont {List}, \citenamefont {Meier}, \citenamefont {Meyer}
  \emph {et~al.}}]{ehret2010new}%
  \BibitemOpen
  \bibfield  {author} {\bibinfo {author} {\bibfnamefont {K.}~\bibnamefont
  {Ehret}}, \bibinfo {author} {\bibfnamefont {M.}~\bibnamefont {Frede}},
  \bibinfo {author} {\bibfnamefont {S.}~\bibnamefont {Ghazaryan}}, \bibinfo
  {author} {\bibfnamefont {M.}~\bibnamefont {Hildebrandt}}, \bibinfo {author}
  {\bibfnamefont {E.-A.}\ \bibnamefont {Knabbe}}, \bibinfo {author}
  {\bibfnamefont {D.}~\bibnamefont {Kracht}}, \bibinfo {author} {\bibfnamefont
  {A.}~\bibnamefont {Lindner}}, \bibinfo {author} {\bibfnamefont
  {J.}~\bibnamefont {List}}, \bibinfo {author} {\bibfnamefont {T.}~\bibnamefont
  {Meier}}, \bibinfo {author} {\bibfnamefont {N.}~\bibnamefont {Meyer}}, \emph
  {et~al.},\ }\bibfield  {title} {\bibinfo {title} {New alps results on
  hidden-sector lightweights},\ }\href@noop {} {\bibfield  {journal} {\bibinfo
  {journal} {Physics Letters B}\ }\textbf {\bibinfo {volume} {689}},\ \bibinfo
  {pages} {149} (\bibinfo {year} {2010})}\BibitemShut {NoStop}%
\bibitem [{\citenamefont {Liu}\ \emph {et~al.}(2019)\citenamefont {Liu},
  \citenamefont {Elwood}, \citenamefont {Evans},\ and\ \citenamefont
  {Thaler}}]{liu2019searching}%
  \BibitemOpen
  \bibfield  {author} {\bibinfo {author} {\bibfnamefont {H.}~\bibnamefont
  {Liu}}, \bibinfo {author} {\bibfnamefont {B.~D.}\ \bibnamefont {Elwood}},
  \bibinfo {author} {\bibfnamefont {M.}~\bibnamefont {Evans}},\ and\ \bibinfo
  {author} {\bibfnamefont {J.}~\bibnamefont {Thaler}},\ }\bibfield  {title}
  {\bibinfo {title} {Searching for axion dark matter with birefringent
  cavities},\ }\href@noop {} {\bibfield  {journal} {\bibinfo  {journal}
  {Physical Review D}\ }\textbf {\bibinfo {volume} {100}},\ \bibinfo {pages}
  {023548} (\bibinfo {year} {2019})}\BibitemShut {NoStop}%
\bibitem [{\citenamefont {Obata}\ \emph {et~al.}(2018)\citenamefont {Obata},
  \citenamefont {Fujita},\ and\ \citenamefont {Michimura}}]{obata2018optical}%
  \BibitemOpen
  \bibfield  {author} {\bibinfo {author} {\bibfnamefont {I.}~\bibnamefont
  {Obata}}, \bibinfo {author} {\bibfnamefont {T.}~\bibnamefont {Fujita}},\ and\
  \bibinfo {author} {\bibfnamefont {Y.}~\bibnamefont {Michimura}},\ }\bibfield
  {title} {\bibinfo {title} {Optical ring cavity search for axion dark
  matter},\ }\href@noop {} {\bibfield  {journal} {\bibinfo  {journal} {Physical
  review letters}\ }\textbf {\bibinfo {volume} {121}},\ \bibinfo {pages}
  {161301} (\bibinfo {year} {2018})}\BibitemShut {NoStop}%
\bibitem [{\citenamefont {Cole}\ \emph {et~al.}(2013)\citenamefont {Cole},
  \citenamefont {Zhang}, \citenamefont {Martin}, \citenamefont {Ye},\ and\
  \citenamefont {Aspelmeyer}}]{cole2013tenfold}%
  \BibitemOpen
  \bibfield  {author} {\bibinfo {author} {\bibfnamefont {G.~D.}\ \bibnamefont
  {Cole}}, \bibinfo {author} {\bibfnamefont {W.}~\bibnamefont {Zhang}},
  \bibinfo {author} {\bibfnamefont {M.~J.}\ \bibnamefont {Martin}}, \bibinfo
  {author} {\bibfnamefont {J.}~\bibnamefont {Ye}},\ and\ \bibinfo {author}
  {\bibfnamefont {M.}~\bibnamefont {Aspelmeyer}},\ }\bibfield  {title}
  {\bibinfo {title} {Tenfold reduction of brownian noise in high-reflectivity
  optical coatings},\ }\href@noop {} {\bibfield  {journal} {\bibinfo  {journal}
  {Nature Photonics}\ }\textbf {\bibinfo {volume} {7}},\ \bibinfo {pages} {644}
  (\bibinfo {year} {2013})}\BibitemShut {NoStop}%
\bibitem [{\citenamefont {Chalermsongsak}\ \emph {et~al.}(2016)\citenamefont
  {Chalermsongsak}, \citenamefont {Hall}, \citenamefont {Cole}, \citenamefont
  {Follman}, \citenamefont {Seifert}, \citenamefont {Arai}, \citenamefont
  {Gustafson}, \citenamefont {Smith}, \citenamefont {Aspelmeyer},\ and\
  \citenamefont {Adhikari}}]{chalermsongsak2016coherent}%
  \BibitemOpen
  \bibfield  {author} {\bibinfo {author} {\bibfnamefont {T.}~\bibnamefont
  {Chalermsongsak}}, \bibinfo {author} {\bibfnamefont {E.~D.}\ \bibnamefont
  {Hall}}, \bibinfo {author} {\bibfnamefont {G.~D.}\ \bibnamefont {Cole}},
  \bibinfo {author} {\bibfnamefont {D.}~\bibnamefont {Follman}}, \bibinfo
  {author} {\bibfnamefont {F.}~\bibnamefont {Seifert}}, \bibinfo {author}
  {\bibfnamefont {K.}~\bibnamefont {Arai}}, \bibinfo {author} {\bibfnamefont
  {E.~K.}\ \bibnamefont {Gustafson}}, \bibinfo {author} {\bibfnamefont {J.~R.}\
  \bibnamefont {Smith}}, \bibinfo {author} {\bibfnamefont {M.}~\bibnamefont
  {Aspelmeyer}},\ and\ \bibinfo {author} {\bibfnamefont {R.~X.}\ \bibnamefont
  {Adhikari}},\ }\bibfield  {title} {\bibinfo {title} {Coherent cancellation of
  photothermal noise in gaas/al0. 92ga0. 08as bragg mirrors},\ }\href@noop {}
  {\bibfield  {journal} {\bibinfo  {journal} {Metrologia}\ }\textbf {\bibinfo
  {volume} {53}},\ \bibinfo {pages} {860} (\bibinfo {year} {2016})}\BibitemShut
  {NoStop}%
\bibitem [{\citenamefont {Kryhin}\ \emph {et~al.}(2023)\citenamefont {Kryhin},
  \citenamefont {Hall},\ and\ \citenamefont
  {Sudhir}}]{kryhin2023thermorefringent}%
  \BibitemOpen
  \bibfield  {author} {\bibinfo {author} {\bibfnamefont {S.}~\bibnamefont
  {Kryhin}}, \bibinfo {author} {\bibfnamefont {E.~D.}\ \bibnamefont {Hall}},\
  and\ \bibinfo {author} {\bibfnamefont {V.}~\bibnamefont {Sudhir}},\
  }\bibfield  {title} {\bibinfo {title} {Thermorefringent noise in crystalline
  optical materials},\ }\href@noop {} {\bibfield  {journal} {\bibinfo
  {journal} {Physical Review D}\ }\textbf {\bibinfo {volume} {107}},\ \bibinfo
  {pages} {022001} (\bibinfo {year} {2023})}\BibitemShut {NoStop}%
\bibitem [{\citenamefont {Kondratiev}\ \emph {et~al.}(2011)\citenamefont
  {Kondratiev}, \citenamefont {Gurkovsky},\ and\ \citenamefont
  {Gorodetsky}}]{kondratiev2011thermal}%
  \BibitemOpen
  \bibfield  {author} {\bibinfo {author} {\bibfnamefont {N.}~\bibnamefont
  {Kondratiev}}, \bibinfo {author} {\bibfnamefont {A.}~\bibnamefont
  {Gurkovsky}},\ and\ \bibinfo {author} {\bibfnamefont {M.}~\bibnamefont
  {Gorodetsky}},\ }\bibfield  {title} {\bibinfo {title} {Thermal noise and
  coating optimization in multilayer dielectric mirrors},\ }\href@noop {}
  {\bibfield  {journal} {\bibinfo  {journal} {Physical Review D}\ }\textbf
  {\bibinfo {volume} {84}},\ \bibinfo {pages} {022001} (\bibinfo {year}
  {2011})}\BibitemShut {NoStop}%
\bibitem [{\citenamefont {Harry}\ \emph {et~al.}(2012)\citenamefont {Harry},
  \citenamefont {Bodiya},\ and\ \citenamefont {DeSalvo}}]{harry2012optical}%
  \BibitemOpen
  \bibfield  {author} {\bibinfo {author} {\bibfnamefont {G.}~\bibnamefont
  {Harry}}, \bibinfo {author} {\bibfnamefont {T.~P.}\ \bibnamefont {Bodiya}},\
  and\ \bibinfo {author} {\bibfnamefont {R.}~\bibnamefont {DeSalvo}},\
  }\href@noop {} {\emph {\bibinfo {title} {Optical coatings and thermal noise
  in precision measurement}}}\ (\bibinfo  {publisher} {Cambridge University
  Press},\ \bibinfo {year} {2012})\BibitemShut {NoStop}%
\bibitem [{\citenamefont {Yariv}\ and\ \citenamefont
  {Yeh}(1983)}]{yariv1983optical}%
  \BibitemOpen
  \bibfield  {author} {\bibinfo {author} {\bibfnamefont {A.}~\bibnamefont
  {Yariv}}\ and\ \bibinfo {author} {\bibfnamefont {P.}~\bibnamefont {Yeh}},\
  }\bibfield  {title} {\bibinfo {title} {Optical waves in crystal propagation
  and control of laser radiation},\ }\href@noop {} {\  (\bibinfo {year}
  {1983})}\BibitemShut {NoStop}%
\bibitem [{\citenamefont {Vyatchanin}(2020)}]{vyatchanin2020loss}%
  \BibitemOpen
  \bibfield  {author} {\bibinfo {author} {\bibfnamefont {S.}~\bibnamefont
  {Vyatchanin}},\ }\bibfield  {title} {\bibinfo {title} {The loss in reflecting
  coating induced by polarization},\ }\href@noop {} {\bibfield  {journal}
  {\bibinfo  {journal} {Physics Letters A}\ }\textbf {\bibinfo {volume}
  {384}},\ \bibinfo {pages} {126878} (\bibinfo {year} {2020})}\BibitemShut
  {NoStop}%
\bibitem [{\citenamefont {Sun}\ \emph {et~al.}(2009)\citenamefont {Sun},
  \citenamefont {Thompson},\ and\ \citenamefont {Nishida}}]{sun2009strain}%
  \BibitemOpen
  \bibfield  {author} {\bibinfo {author} {\bibfnamefont {Y.}~\bibnamefont
  {Sun}}, \bibinfo {author} {\bibfnamefont {S.~E.}\ \bibnamefont {Thompson}},\
  and\ \bibinfo {author} {\bibfnamefont {T.}~\bibnamefont {Nishida}},\
  }\href@noop {} {\emph {\bibinfo {title} {Strain effect in semiconductors:
  theory and device applications}}}\ (\bibinfo  {publisher} {Springer Science
  \& Business Media},\ \bibinfo {year} {2009})\BibitemShut {NoStop}%
\bibitem [{\citenamefont {Adachi}(1993)}]{adachi1993properties}%
  \BibitemOpen
  \bibfield  {author} {\bibinfo {author} {\bibfnamefont {S.}~\bibnamefont
  {Adachi}},\ }\href@noop {} {\emph {\bibinfo {title} {Properties of aluminium
  gallium arsenide}}},\ \bibinfo {number} {7}\ (\bibinfo  {publisher} {IET},\
  \bibinfo {year} {1993})\BibitemShut {NoStop}%
\bibitem [{\citenamefont {Afromowitz}(1974)}]{afromowitz1974refractive}%
  \BibitemOpen
  \bibfield  {author} {\bibinfo {author} {\bibfnamefont {M.~A.}\ \bibnamefont
  {Afromowitz}},\ }\bibfield  {title} {\bibinfo {title} {Refractive index of
  ga1- xalxas},\ }\href@noop {} {\bibfield  {journal} {\bibinfo  {journal}
  {Solid State Communications}\ }\textbf {\bibinfo {volume} {15}},\ \bibinfo
  {pages} {59} (\bibinfo {year} {1974})}\BibitemShut {NoStop}%
\bibitem [{\citenamefont {Cole}\ \emph {et~al.}(2011)\citenamefont {Cole},
  \citenamefont {Wilson-Rae}, \citenamefont {Werbach}, \citenamefont {Vanner},\
  and\ \citenamefont {Aspelmeyer}}]{cole2011phonon}%
  \BibitemOpen
  \bibfield  {author} {\bibinfo {author} {\bibfnamefont {G.~D.}\ \bibnamefont
  {Cole}}, \bibinfo {author} {\bibfnamefont {I.}~\bibnamefont {Wilson-Rae}},
  \bibinfo {author} {\bibfnamefont {K.}~\bibnamefont {Werbach}}, \bibinfo
  {author} {\bibfnamefont {M.~R.}\ \bibnamefont {Vanner}},\ and\ \bibinfo
  {author} {\bibfnamefont {M.}~\bibnamefont {Aspelmeyer}},\ }\bibfield  {title}
  {\bibinfo {title} {Phonon-tunnelling dissipation in mechanical resonators},\
  }\href@noop {} {\bibfield  {journal} {\bibinfo  {journal} {Nature
  communications}\ }\textbf {\bibinfo {volume} {2}},\ \bibinfo {pages} {231}
  (\bibinfo {year} {2011})}\BibitemShut {NoStop}%
\bibitem [{\citenamefont {Dixon}(1967)}]{dixon1967photoelastic}%
  \BibitemOpen
  \bibfield  {author} {\bibinfo {author} {\bibfnamefont {R.}~\bibnamefont
  {Dixon}},\ }\bibfield  {title} {\bibinfo {title} {Photoelastic properties of
  selected materials and their relevance for applications to acoustic light
  modulators and scanners},\ }\href@noop {} {\bibfield  {journal} {\bibinfo
  {journal} {Journal of Applied Physics}\ }\textbf {\bibinfo {volume} {38}},\
  \bibinfo {pages} {5149} (\bibinfo {year} {1967})}\BibitemShut {NoStop}%
\bibitem [{\citenamefont {Detraux}\ and\ \citenamefont
  {Gonze}(2001)}]{detraux2001photoelasticity}%
  \BibitemOpen
  \bibfield  {author} {\bibinfo {author} {\bibfnamefont {F.}~\bibnamefont
  {Detraux}}\ and\ \bibinfo {author} {\bibfnamefont {X.}~\bibnamefont
  {Gonze}},\ }\bibfield  {title} {\bibinfo {title} {Photoelasticity of
  $\alpha$-quartz from first principles},\ }\href@noop {} {\bibfield  {journal}
  {\bibinfo  {journal} {Physical Review B}\ }\textbf {\bibinfo {volume} {63}},\
  \bibinfo {pages} {115118} (\bibinfo {year} {2001})}\BibitemShut {NoStop}%
\bibitem [{\citenamefont {Donadio}\ \emph {et~al.}(2003)\citenamefont
  {Donadio}, \citenamefont {Bernasconi},\ and\ \citenamefont
  {Tassone}}]{donadio2003photoelasticity}%
  \BibitemOpen
  \bibfield  {author} {\bibinfo {author} {\bibfnamefont {D.}~\bibnamefont
  {Donadio}}, \bibinfo {author} {\bibfnamefont {M.}~\bibnamefont
  {Bernasconi}},\ and\ \bibinfo {author} {\bibfnamefont {F.}~\bibnamefont
  {Tassone}},\ }\bibfield  {title} {\bibinfo {title} {Photoelasticity of
  crystalline and amorphous silica from first principles},\ }\href@noop {}
  {\bibfield  {journal} {\bibinfo  {journal} {Physical Review B}\ }\textbf
  {\bibinfo {volume} {68}},\ \bibinfo {pages} {134202} (\bibinfo {year}
  {2003})}\BibitemShut {NoStop}%
\end{thebibliography}%

\end{document}